\documentclass[aps,nofootinbib]{revtex4-2}

\usepackage{graphicx,epsfig}
\usepackage{amsmath}
\usepackage{subfigure}
\usepackage{dcolumn}
\usepackage{amsmath}
\usepackage{amsfonts}
\usepackage{amssymb}
\usepackage{braket}
\usepackage{amssymb}
\usepackage{amsmath}
\usepackage{graphicx}
\usepackage{subfigure}
\usepackage{makeidx}
\usepackage{capt-of}
\usepackage{float}
\usepackage{placeins}
\usepackage{perpage}
\usepackage{mwe}
\usepackage{kantlipsum}
\usepackage{graphicx}
\usepackage{amsfonts}
\usepackage{xcolor}
\begin{document}

\title{Constant-roll inflation driven by holographic dark energy}
\author{Abolhassan Mohammadi}
   \email{a.mohammadi@uok.ac.ir; abolhassanm@gmail.com}
   \affiliation{Department of Physics, Faculty of Science, University of Kurdistan,
Pasdaran Street, P.O. Box 66177-15175 Sanandaj, Iran.}
%    \affiliation{Physics}
%\author{Mr Physics}
%   \email{ggg@gmail.com}
%   \affiliation{Department}

\begin{abstract}
The scenario of constant-roll inflation is investigated assuming that it is caused by holographic dark energy with energy density $3M_p^2 c^2 / L^2$. The energy density with varying $c$ parameter and Hubble radius as infrared cutoff provides a proper description for the late time acceleration expansion. This motivates us to study the role of the same holographic dark energy in the scenario of constant-roll inflation.  One of the main features of constant-roll inflation is that the second slow-roll parameter is constant which is not necessarily small. This feature brings a differential equation for the parameter $c$. After finding an analytical solution for the parameter, the observable parameters are obtained at the time of inflation, and by comparing the model with data, the constants of the model are determined. The energy scale of inflation is estimated which is about $10^{16} \; {\rm GeV}$. Finally, a correspondence between the HDE and canonical scalar field is considered and the behavior of the constructed potential and its magnitude is discussed.

\end{abstract}

\date{\today}
\maketitle
%\tableofcontents

%%%%%%%%%%%%%%%%%%%%%%%%%%%%%%%%%%%%%%%%%%
%%%%%%%%%%%%%%%%%%%%%%%%%%%%%%%%%%%%%%%%%%
%%%%%%%%%%%%%%%%%%%%%%%%%%%%%%%%%%%%%%%%%%
%%%%%%%%%%%%%%%%%%%%%%%%%%%%%%%%%%%%%%%%%%\\
%%%%%%%%%%%%%%%%%%%%%%%%%%%%%%%%%%%%%%%%%%
%%%%%%%%%%%%%%%%%%%%%%%%%%%%%%%%%%%%%%%%%%
%%%%%%%%%%%%%%%%%%%%%%%%%%%%%%%%%%%%%%%%%%
%%%%%%%%%%%%%%%%%%%%%%%%%%%%%%%%%%%%%%%%%%
%%%%%%%%%%%%%%%%%%%%%%%%%%%%%%%%%%%%%%%%%%
\section{Introduction}\label{Sec_intro}
Inflation is known as one of the candidate scenarios for describing the universe evolution at very early times. It states that the universe undergoes an extremely accelerated expansion in a very short time. Such a theory of expansion for the early times was first introduced by Starobinsky \cite{starobinsky1980new}, and then it was first utilized by Guth \cite{Guth:1980zm} as a possible solution for the problems of the hot big-bang theory. Due to the great consistency with data \cite{Planck:2018jri}, the scenario has been widely studied, and generalized in many different ways \cite{Barenboim:2007ii,Franche:2010yj,Unnikrishnan:2012zu,Rezazadeh:2014fwa,Saaidi:2015kaa,Fairbairn:2002yp,Mukohyama:2002cn,Feinstein:2002aj,Padmanabhan:2002cp,Aghamohammadi:2014aca,Spalinski:2007dv,Bessada:2009pe,Weller:2011ey,Nazavari:2016yaa,maeda2013stability,abolhasani2014primordial,alexander2015dynamics,tirandari2017anisotropic,maartens2000chaotic,golanbari2014brane,Mohammadi:2020ake,Mohammadi:2020ctd,berera1995warm,berera2000warm,hall2004scalar,Sayar:2017pam,Akhtari:2017mxc,Sheikhahmadi:2019gzs,Rasheed:2020syk,Mohammadi:2018oku,Mohammadi:2019dpu,Mohammadi:2018zkf,Mohammadi:2019qeu,Mohammadi:2020ftb}. \\

In most inflationary models, inflation is driven by a scalar field, called inflaton, which dominates the universe. Moreover, it is assumed that the scalar field slowly rolls down from the top of its potential toward the minimum. Such approximation is known as the slow-roll approximations, and the model built based on it is addressed as the slow-roll inflation. This behavior is well expressed based on the slow-roll parameters which should be small during inflationary times. The smallness of these parameters is usually provided by an almost flat potential \cite{Linde:2000kn,Linde:2005ht,Linde:2005vy,Linde:2004kg,Riotto:2002yw,Baumann:2009ds,Weinberg:2008zzc,Lyth:2009zz}. \\
What happens if the potential is exactly flat? in this case, the second slow-roll parameter becomes constant and of the order of one; instead of being small.  The question was first studied in \cite{Kinney:2005vj}. Then, the idea was followed by \cite{Martin:2012pe,Motohashi:2014ppa}, and it led to the introduction of constant-roll inflation. In constant-roll inflation, the second slow-roll parameter is taken as a constant which in general might be of the order of one. Utilizing Hamilton-Jacobi formalism \cite{Salopek:1992qy,Liddle:1994dx,Kinney:1997ne,Guo:2003zf,Aghamohammadi:2014aca,Saaidi:2015kaa,
Sheikhahmadi:2016wyz,Rasheed:2020syk}, it is possible to find an exact solution for the model. \\

The goal we are going to pursue is considering constant-roll inflation with the assumption that it is the holographic dark energy (HDE) that drives inflation. The HDE is known as one of the candidates of dark energy \cite{Hsu:2004ri,Horvat:2004vn,Li:2004rb} which has been widely used in the late time evolution of the universe. It is built based on the holographic principle \cite{tHooft:1993dmi} stating that the entropy of a system is scaled based on it surface area \cite{tHooft:1993dmi,Susskind:1994vu,Witten:1998qj,Bousso:2002ju}. Inspiring by the thermodynamics of the black hole, the use of HDE in cosmological studies received huge interest, and it also is extended to the string theory \cite{Susskind:1994vu}. \\
The original HDE is based on the standard entropy relation $S = A/4$ of the black hole where $A$ stands for the area of the horizon. Including the quantum corrections to the entropy, there are different modifications for the entropy such as the logarithmic entropy \cite{Mann:1996ze,Rovelli:1996dv,Ashtekar:1997yu,Kaul:2000kf,Das:2001ic} arises from loop quantum gravity and power-law correction is due to the entanglement of quantum field \cite{Das:2008sy,Das:2010su,Das:2007mj,Radicella:2010ss}. \\

The applications of HDE and its modified versions in late-time cosmology and dark energy have been extensively considered \cite{Nojiri:2019skr,Nojiri:2019itp,Nojiri:2005pu,Nojiri:2020wmh} (for a review on HDE refer to \cite{Li:2004rb}). Recently, the role of HDE in early times has received interest and it was picked as the topic of some research proposals \cite{Nojiri:2019kkp,Oliveros:2019rnq,Chakraborty_2020,Mohammadi:2021wde}. However, there has not been any effort for investigating the HDE as a source of constant-roll inflation. Therefore, it would be a good inspiration to study the constant-roll inflation where the HDE is playing the role of inflation energy density.   \\

The paper is organized as follows: In Sec.\ref{inflation_basic} the basic of inflation is presented, and we briefly introduced constant-roll inflation. Applying HDE in constant-roll inflation is discussed in Sec.\ref{HDE_crinflation}. It is described that the constancy of the second slow-roll parameter leads to a differential equation for the cutoff and a solution for the cutoff is found out. Then, the results are compared with the observational data and a range for the constants of the model is achieved. The results of the model are summarized in Sec.\ref{conclusion}.

%%%%%%%%%%%%%%%%%%%%%%%%%%%%%%%%%%%%%%%%%%
%%%%%%%%%%%%%%%%%%%%%%%%%%%%%%%%%%%%%%%%%%
%%%%%%%%%%%%%%%%%%%%%%%%%%%%%%%%%%%%%%%%%%
%%%%%%%%%%%%%%%%%%%%%%%%%%%%%%%%%%%%%%%%%%\\
%%%%%%%%%%%%%%%%%%%%%%%%%%%%%%%%%%%%%%%%%%
%%%%%%%%%%%%%%%%%%%%%%%%%%%%%%%%%%%%%%%%%%
%%%%%%%%%%%%%%%%%%%%%%%%%%%%%%%%%%%%%%%%%%
%%%%%%%%%%%%%%%%%%%%%%%%%%%%%%%%%%%%%%%%%%
%%%%%%%%%%%%%%%%%%%%%%%%%%%%%%%%%%%%%%%%%%
\section{Inflation: Basics}\label{inflation_basic}
The dynamics of the universe is usually described by the Friedmann equations
\begin{eqnarray}\label{Friedmann_full}
H^2 & = & {1 \over 3M_p^2} \; \rho_t  , \\
\dot{H} & = & {-1 \over 2 M_p^2} \; (\rho_t + p_t).
\end{eqnarray}
Here $\rho_t$ is the total energy density of the universe, which in general is a combination of different fluids, and $p_t$ is the total pressure. The conservation equation for the total energy density is given by
\begin{equation}\label{conservation_full}
\dot{\rho}_t + 3 H (\rho_t + p_t) = 0.
\end{equation}

Based on the inflationary scenario, it is assumed that the universe is dominated by the scalar field with the following energy density and pressure
\begin{eqnarray}\label{sf_energypressure}
\rho_\phi & = & {1 \over 2} \; \dot{\phi}^2 + V(\phi), \\
p_\phi & = & {1 \over 2} \; \dot{\phi}^2 - V(\phi).
\end{eqnarray}
Substituting the above energy density and pressure in Eq.\eqref{conservation_full}, one arrives at
\begin{equation}\label{sf_eom}
\ddot{\phi} + 3H \dot{\phi} + V'(\phi) = 0,
\end{equation}
where the prime indicates derivative with respect to the scalar field, and the equation also is known as the equation of motion of the scalar field. \\
Therefore, during the inflationary time, the evolution of the universe is governed by the scalar field which produces an accelerated phase.

%Some parameters play an important role in inflationary models, which are known as slow-roll parameters. The first slow-roll parameter is defined as the rate of the Hubble parameter during a Hubble time, i.e.
%\begin{equation}\label{epsilon1_full}
%\epsilon = {-\dot{H} \over H^2}.
%\end{equation}
%The other slow-roll parameters are defined by the following equation
%\begin{equation}\label{srp_all}
%\epsilon_{n+1} = {\dot{\epsilon}_n \over H \epsilon_n},
%\end{equation}
%for $n \geq 0$, where $\epsilon_1$ is given by Eq.\eqref{epsilon1_full}. The second slow-roll parameter is given by
%\begin{equation}\label{epsilon2}
%\epsilon_2 = {\dot{\epsilon}_1 \over H \epsilon_1}.
%\end{equation}

%%%%%%%%%%%%%%%%%%%%%%%%%%%%%%%%%%%%%%%%%%\\
%%%%%%%%%%%%%%%%%%%%%%%%%%%%%%%%%%%%%%%%%%
%%%%%%%%%%%%%%%%%%%%%%%%%%%%%%%%%%%%%%%%%%
%%%%%%%%%%%%%%%%%%%%%%%%%%%%%%%%%%%%%%%%%%
%%%%%%%%%%%%%%%%%%%%%%%%%%%%%%%%%%%%%%%%%%
%%%%%%%%%%%%%%%%%%%%%%%%%%%%%%%%%%%%%%%%%%
\subsection{Slow-roll inflation}\label{srinflation}
The first model which gives an accelerated expansion is the de-Sitter model. However, the problem with the de-Sitter model is that the Hubble parameter is constant and the universe could not exit from the accelerated expansion phase. One solution for the problem is to assume a quasi-de Sitter expansion instead of de-Sitter expansion. To have a quasi-de Sitter expansion, the rate of the Hubble parameter during a Hubble time should be very small \cite{Riotto:2002yw,Baumann:2009ds,Weinberg:2008zzc,Lyth:2009zz}, i.e.
\begin{equation}\label{epsilon1_full}
\epsilon_1 = {-\dot{H} \over H^2} \ll 1
\end{equation}
in which $\epsilon_1$ is called the first slow-roll parameter. Utilizing Eqs.\eqref{Friedmann_full} and \eqref{sf_energypressure}, it leads to this conclusion that to have a quasi-de Sitter expansion, the kinetic energy of the scalar field should be smaller than the its potential \cite{Riotto:2002yw,Baumann:2009ds,Weinberg:2008zzc,Lyth:2009zz}. The same argument is applied for $\dot\phi$, so that the rate of $\dot{\phi}$ during a Hubble time is assumed to be smaller than unity, \cite{Weinberg:2008zzc}
\begin{equation}\label{eta}
\delta = {-\ddot{\phi} \over H \dot{\phi} } \ll 1,
\end{equation}
where $\delta$ is known as the second slow-roll parameter. \\
In the slow-roll inflation, the slow-roll parameters are smaller than one during the inflationary times. Applying these conditions on the Friedmann equations \eqref{Friedmann_full} and scalar field equation of motion \eqref{sf_eom}, leads to
\begin{eqnarray}\label{srinflation_equations}
H^2 & \simeq & {1 \over 3M_p^2} \; V(\phi), \\
\dot{H} & = & {-1 \over 2M_p^2} \; \dot{\phi}^2, \\
3H \dot{\phi} & \simeq & - V'(\phi).
\end{eqnarray}
Before going further, it is worth mentioning that, there is another way to arrive at parameter $\delta$ in Eq.\eqref{eta}. There is a hierarchy relation to define a consecutive slow-roll parameter as $\epsilon_{n+1} = \dot{\epsilon}_n / H \epsilon_n$ for $n \geq 1$. For the second parameter, one arrives at $\epsilon_2 = 2 \epsilon_1 + 2\ddot{\phi}/H\dot{\phi} $. Since, both $\epsilon_1 $ and $\epsilon_2$ are slow-roll parameter and smaller than one, the second term on r.h.s also should be small. Then, the term $\ddot{\phi} / H\dot{\phi}$ is addressed as a slow-roll parameter as well. \\

Applying Eqs.\eqref{srinflation_equations}, the slow-roll parameters $\epsilon_1$ and $\eta$ could be expressed in terms of the potential and its derivatives as \cite{Riotto:2002yw,Baumann:2009ds,Weinberg:2008zzc,Lyth:2009zz}
\begin{equation}
\epsilon_1 = {M_p^2 \over 2} \; {V'^2 \over V^2}, \quad {\rm and} \quad \eta = M_p^2 \; {V'' \over V} .
\end{equation}
That is the reason why it is stated that the smallness of the slow-roll parameters is supported by an almost flat potential. To investigate the model in more detail, and compare it with the observational data, it is required to introduce a potential for the model. So far, the slow-roll inflation has been considered for many different types of potential; refer to \cite{Baumann:2009ds} for some of these potentials.

%%%%%%%%%%%%%%%%%%%%%%%%%%%%%%%%%%%%%%%%%%\\
%%%%%%%%%%%%%%%%%%%%%%%%%%%%%%%%%%%%%%%%%%
%%%%%%%%%%%%%%%%%%%%%%%%%%%%%%%%%%%%%%%%%%
%%%%%%%%%%%%%%%%%%%%%%%%%%%%%%%%%%%%%%%%%%
%%%%%%%%%%%%%%%%%%%%%%%%%%%%%%%%%%%%%%%%%%
%%%%%%%%%%%%%%%%%%%%%%%%%%%%%%%%%%%%%%%%%%
\subsection{Constant-roll inflation}
In the case of flat potential, there is $V' = 0$, and according to Eq.\eqref{sf_eom}, one has $\ddot{\phi} = -3 H \dot\phi$. Then, the slow-roll parameter $\eta = 3$ \cite{Kinney:2005vj}. It is no longer small. This was the start point of the scenario of constant-roll inflation, where the smallness of the second slow-roll parameter was released. \\
Constant-roll inflation takes one step beyond slow-roll inflation. Here, the second slow-roll parameter is taken as a constant which in general could be of the order of unity. One of the main features of constant-roll inflation is that there is no longer a need to introduce a random potential. The constancy of the second slow-roll parameter leads to a differential equation for the Hubble parameter, and consequently a solution for the model 
\cite{Motohashi:2014ppa,Motohashi:2017aob,Motohashi:2019tyj,Motohashi:2019rhu}. \\
The scenario was studied in \cite{Motohashi:2014ppa}. The authors take the second slow-roll parameter $\eta$ as
\begin{equation}\label{cr_2srp}
\eta = {\ddot{\phi} \over H \dot{\phi}} = (3 + \alpha) ,
\end{equation}
where $\alpha$ is a constant. Applying the Hamilton-Jacobi formalism\footnote{In the Hamilton-Jacobi formalism, the Hubble parameter is assumed to be a function of the scalar field.}, and by using Eqs.\eqref{srinflation_equations}, one obtains the differential equation for the Hubble parameter, as follows
\begin{equation}\label{cr_HubbleDE}
{d^2 H(\phi) \over d\phi^2} = {3+\alpha \over 2M_p^2} \; H.
\end{equation}
Depending on the value of the constant $\alpha$, there are three types of solution for the $H(\phi)$. It was realized that the best solution is for $\alpha < -3$, and for this case, the model comes to a good consistency with observational data.

%%%%%%%%%%%%%%%%%%%%%%%%%%%%%%%%%%%%%%%%%%
%%%%%%%%%%%%%%%%%%%%%%%%%%%%%%%%%%%%%%%%%%
%%%%%%%%%%%%%%%%%%%%%%%%%%%%%%%%%%%%%%%%%%
%%%%%%%%%%%%%%%%%%%%%%%%%%%%%%%%%%%%%%%%%%\\
%%%%%%%%%%%%%%%%%%%%%%%%%%%%%%%%%%%%%%%%%%
%%%%%%%%%%%%%%%%%%%%%%%%%%%%%%%%%%%%%%%%%%
%%%%%%%%%%%%%%%%%%%%%%%%%%%%%%%%%%%%%%%%%%
%%%%%%%%%%%%%%%%%%%%%%%%%%%%%%%%%%%%%%%%%%
%%%%%%%%%%%%%%%%%%%%%%%%%%%%%%%%%%%%%%%%%%
\section{HDE for constant-roll inflation}\label{HDE_crinflation}
The acceleration phase of the universe during inflation is provided by dark energy which is usually assumed to be a scalar field. Here, we are going to assume that it is the HDE that provides the necessary energy for inflation. \\
The original HDE is given by
\begin{equation}\label{OHDE}
\rho = {3 c^2 M_p^2 \over L^2},
\end{equation}
where $c$ is a dimensionless parameter, and $L$ is known as the infrared cutoff.
In the cosmological studies the infrared cutoff is taken to be a form of the horizon, such as Hubble horizon or particle horizon. However, Only due to the dimensional motivation, there are other choices for the cutoff as Ricci scalar $L^{-2} = R = H^2 + \dot{H}$, or the Granda-Oliveros cutoff $L^{-2} = \alpha H^2 + \beta \dot{H}$ \cite{Granda:2008dk,Granda:2008tm}, which could be count as a modified version of Ricci scalar. \\
The HDE with Hubble length as the infrared cutoff is not a suitable case for describing the current expansion of the universe. However, in a more general viewpoint, the parameter $c$ is assumed as a varying function \cite{Radicella_2010,Pavon:2005yx,Pav_n_2007,Guberina_2007}. In \cite{Malekjani_prd_2018}, it was shown that for the case of varying parameter $c$, the model have a good consistency with data, and provides a proper explanation about the universe expansion. Motivated by this result, it is aimed to consider the constant-roll inflation where the HDE, with Hubble length and varying $c$-term, plays the source role. Since it is believed that inflation occurs in a high energy regime, a correction from the Ultraviolet cutoff is taken for the infrared cutoff as $L^2 \rightarrow L^2 + {1 \over \Lambda^2_{UV}}$ \cite{Nojiri:2019kkp,Oliveros:2019rnq,Chakraborty_2020}. Then, the length scale we are working with is given by $L^2 = 1/H^2 + 1/\Lambda^2_{UV}$, and the HDE is read as
\begin{equation}\label{HDE_UVcorection}
  \rho = {3 M_p^2 c^2 \over {{1 \over H^2} + {1 \over \Lambda^2_{UV}}}}.
\end{equation}
Substituting Eq.\eqref{HDE_UVcorection} in the Eq.\eqref{Friedmann_full}, the Hubble parameter is obtained
\begin{equation}\label{HDE_Friedmann}
H^2 = \Lambda^2_{UV} \; \big( c^2 - 1 \big).
\end{equation}
Then, using the definition \eqref{epsilon1_full}, the first slow-roll parameter $\epsilon_1$ is expressed as
\begin{equation}\label{HDE_epsilon1}
\epsilon_1 = {-1 \over \Lambda_{UV}} \; {c \; \dot{c} \over \big( c^2 - 1 \big)^{3/2}}
\end{equation}
From the definition of $\epsilon_2$, one arrives at
\begin{equation}
\epsilon_2 = {1 \over \Lambda \sqrt{c^2 - 1} } \; \left( {\dot{c} \over c} + {\ddot{c} \over \dot{c}} \right) 
                    -  {3 \over \Lambda} \; {c \; \dot{c} \over \big( c^2 - 1 \big)^{3/2}} 
                    =  {1 \over \Lambda \sqrt{c^2 - 1} } \; \left( {\dot{c} \over c} + {\ddot{c} \over \dot{c}} \right) 
                    + 3 \epsilon_1
\end{equation}  
where Eq.\eqref{HDE_epsilon1} has been used. From above relation, and by following the same procedure as explained in Sec.\ref{srinflation}, the second slow-roll parameter $\eta$ is defined through the parameter $\epsilon_2$ as 
\begin{equation}
\eta = {1 \over \Lambda \sqrt{c^2 - 1} } \; \left( {\dot{c} \over c} + {\ddot{c} \over \dot{c}} \right).
\end{equation} 
therefore we have $\epsilon_2 = \eta + 3 \epsilon_1$.  \\
In constant-roll inflation, the second slow-roll parameter is taken as a constant, i.e. $\eta = \beta$. Then, Eq.\eqref{HDE_cr2srp} gives a differential equation for the parameter $c$. Integrating leads to the following solution
\begin{equation}\label{ccdot_solution}
c \; \dot{c} = {\beta \; \Lambda_{UV} \over 3} \; \Big( (c^2 - 1)^{3/2} + d_0 \Big),
\end{equation}
where $d_0$ is a constant of integration. Using this result, the first slow-roll parameter is rewritten as
\begin{equation}\label{epsilon1_vs_c}
\epsilon_1 = {-\beta \over 3} \; {\big( c^2 - 1 \big)^{3/2} + d_0 \over \big( c^2 - 1 \big)^{3/2}}.
\end{equation}
The slow-roll parameter $\epsilon_1$ reaches one at the end of inflation, which indicates the end of accelerated expansion phase. Then, through the relation $\epsilon_1 = 1$, the parameter $c$ at the end of inflationary phase is obtained as
\begin{equation}\label{c_end}
\big( c^2_e - 1 \big)^{3/2} = {- \beta \; d_0 \over 3 + \beta},
\end{equation}
where the subscript 'e' means the end of inflation. By applying the number of e-fold equation,
\begin{equation} \nonumber
N = \int_{t_\star}^{t_e} \; H \; dt,
\end{equation}
and using Eq.\eqref{c_end}, the parameter $c$ will be acquired in terms of the number of e-fold
\begin{equation}\label{c_vs_N}
\big( c^2_\star - 1 \big)^{3/2} =  d_0 \; \left( {3 \over 3 + \beta} \; e^{-\beta N} - 1 \right),
\end{equation}
in which the subscript '$\star$' stands for the parameter during inflation. Substituting Eq.\eqref{c_vs_N} in Eq.\eqref{epsilon1_vs_c}, the first slow-roll parameter is read in terms of the number of e-fold
\begin{equation}\label{epsilon1_vs_N}
\epsilon_1 = {-\beta \over 3} \; {e^{-\beta N} \over e^{-\beta N} - \left( 1 + {\beta \over 3} \right)}.
\end{equation}

\begin{figure}[h]
\centering
\includegraphics[width=8cm]{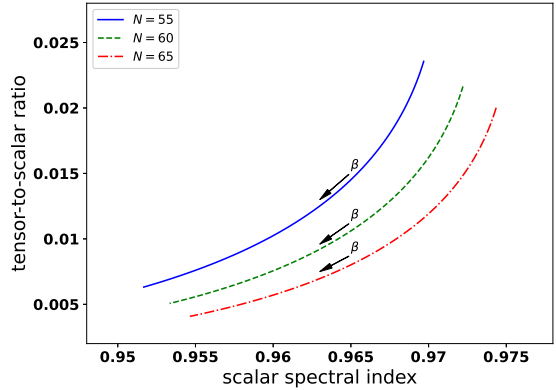}
\caption{The figure shows the tensor-to-scalar ratio versus the scalar spectral index for different values of the number of e-fold. The parameter $\beta$ is varying parameter and it increase along the direction of the arrows.}\label{rns}
\end{figure}

\begin{figure}[ht]
\centering
\includegraphics[width=8cm]{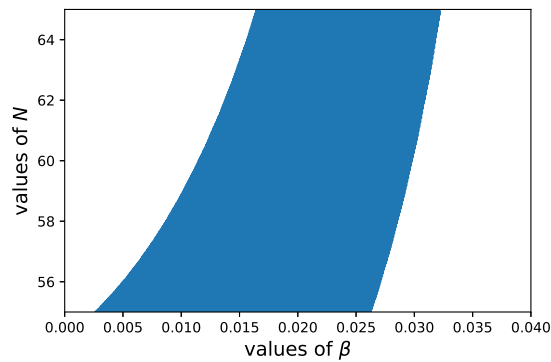}
\caption{\label{beta_N_space} The figure illustrates the set of $(\beta, N)$ pair as a parametric space. For each $(\beta, N)$ point in this parametric, the model gives $n_r$ and $r$ exactly in the data range. Therefore, the space shows a set of $(\beta, N)$ that brings the model to an agreement with data. }
\end{figure}

An inflationary model is required to be compared with observational data to be reliable. In this regard, the model estimation about the perturbation parameters should be calculated, in which the amplitude of the scalar perturbations, scalar spectral index, and the tensor-to-scalar ratio are our main interest. \\
The observational parameters of the constant-roll inflation are discussed in great detail in \cite{Motohashi:2014ppa}. These parameters are obtained as
\begin{eqnarray}
n_s & = & 2 - 4\nu^2, \label{cr_ns} \\
\nu^2 & = & {9 \over 4} +  3\epsilon_1 + {3 \over 2} \; \epsilon_2 + {5 \over 2} \; \epsilon_1 \epsilon_2 + {1 \over 2} \; \epsilon_2 \epsilon_3  \nonumber \\
      & &  \qquad \qquad + {1 \over 4} \; \epsilon_2^2 + {1 \over 2} \; \epsilon_1 \epsilon_2^2  \label{cr_nu} \\
r & = & 16 \left( \Gamma(3/2) \over 2^{\nu - {1 \over 2}} \; \Gamma(\nu) \right)^2 \; \epsilon_1 . \label{cr_r}
\end{eqnarray}
Note that $\epsilon_3$ is another slow-roll parameter defined as $\epsilon_3 = \dot{\epsilon}_2 / H \epsilon_2$ and $\epsilon_2 = \eta + 3\epsilon_1$. Doing a little calculation, it is proved that $\epsilon_3 = 3 \epsilon_1$. The amplitude of the scalar perturbation is also given by
\begin{equation}\label{cr_ps}
\mathcal{P}_s = \left( 2^{\nu - 1/2} \Gamma(\nu) \over \Gamma(3/2) \right)^2 \;
                      {H^2 \over 8\pi^2 M_p^2 \epsilon_1}.
\end{equation}
Although an exact perturbation investigation for HDE is required, it is assumed that Eqs.\eqref{cr_ns}, \eqref{cr_r}, and \eqref{cr_ps} provide a good approximation about the perturbation parameters \cite{Nojiri:2019kkp,Oliveros:2019rnq,Chakraborty_2020,Mohammadi:2021wde}. Utilizing Eqs.\eqref{HDE_Friedmann}, \eqref{c_vs_N}, and \eqref{epsilon1_vs_N}, the perturbation parameters are obtained in terms of the number of e-fold. Then, they could be estimated at any time of inflation. We are mostly interested to measure the perturbation parameters at the time of the horizon crossing, where the number of e-fold commonly is believed to be about $N=55-65$ \cite{Riotto:2002yw,Baumann:2009ds}.  \\
Substituting Eqs.\eqref{epsilon1_vs_N}, and \eqref{cr_nu} in Eqs.\eqref{cr_ns} and \eqref{cr_r}, the scalar spectral index and the tensor-to-scalar ratio are acquired in terms of the constants $\beta$ and number of e-fold. Fig.\ref{rns} displays the tensor-to-scalar ratio versus the scalar spectral index at the horizon crossing time for three different choices of the number of e-fold. Here, the varying parameter is $\beta$ and it increase is the direction of the arrow, drawn along the curves.
%%%%%%%%%%%%%%%%%%%%%%%%%%
%%%%%%%%%%%%%%%%%%%%%%%%%%%
According to the latest data, the scalar spectral index is about $n_s  = 0.9649 \pm 0.0042$ and the tensor-to-scalar ratio is confined by an upper bound as $r < 0.036$ \cite{BICEP:2021xfz}. It is realized that the $r-n_s$ curves in
Fig.\ref{rns} perfectly cross this interested region indicating the compatibility of the model with data.  It also should be noted that for each specific value of number of e-fold, there is a different consistent value for $\beta$. Using Python coding, one could find a set of $(\beta, N)$ pairs that brings the model to an agreement with data.
Fig.\ref{beta_N_space} portrays this set as a parametric space. For each $(\beta, N)$ point in this space, the estimated value for $n_s$ and $r$ is consistent with data. Also, it is realized that for each specifi value of $N$, there is a range for $\beta$. This range gets smaller as $N$ grows. \\
%%%%%%%%%%%%%%%%%%%%%%%%%%
%%%%%%%%%%%%%%%%%%%%%%%%%%

\begin{figure}
\centering
\includegraphics[width=8cm]{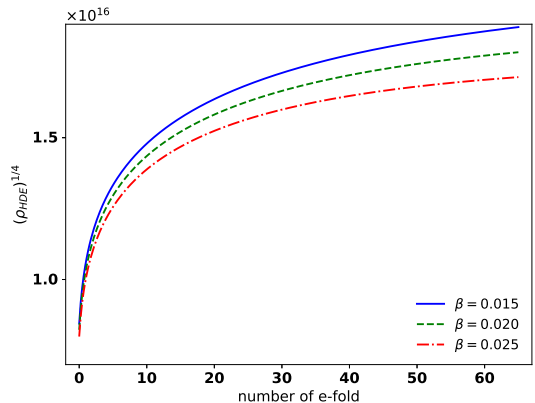}
\caption{The figure shows the energy scale of inflation versus the number of e-fold for $d_0 = -0.1$ and different values of the constant $\beta$. }\label{energy_scale_inflation}
\end{figure}

The amplitude of the scalar perturbations is predicted as $\ln\big( 10^{10} \; \mathcal{P}_s \big) = 3.044 \pm 0.014$ \cite{Planck:2018jri}. This data could be used to determine the constant $\Lambda_{UV}$. Through Eq.\eqref{cr_ps}, the constant is obtained to be
\begin{equation}\label{Lambda}
\Lambda^2_{UV} = \left( {\Gamma(3/2) \over 2^{\nu - 1/2} \Gamma(\nu)}   \right)^2 \;
                               {\big( 8 \pi^2 M_p^2 \; \epsilon_1 \; \mathcal{P}_s \big) \over
                               \left[ d_0 \left( {e^{-\beta N} \over 1 + {\beta \over 3}} + 1 \right) \right]^{2/3}} \;
\end{equation}
The appearance of the constant $d_0$ implies that the model is not fully determined yet. There are two unknown constant $\Lambda_{UV}$ and $d_0$ and one data for the amplitude of the scalar perturbations. However, the mentioned constants also appear in the energy scale equation. The energy scale of inflation is of much interest to the cosmologist because it determines whether the topological defect formation could occur after inflation. It is very important for studying structure formation. The energy scale of inflation is also essential for understanding early Universe physics. The energy scale of inflation, where the fluctuations cross the horizon, is estimated to be a few orders of magnitude smaller than the Planck energy scale, namely about $10^{15}-10^{16} \; {\rm GeV}$ based on the standard model of inflation and the data about the tensor-to-scalar ratio. Although this result is obtained mainly in the standard inflationary model, many other studies agree with this conclusion. In this work, we take this result for the energy scale as another tool to constrain the constant of the model. Doing so, we have two undetermined constants $\Lambda$ and $d_0$, and two equations.   \\
The energy density is presented by $\rho_H$. Substituting Eqs.\eqref{HDE_Friedmann}, \eqref{c_vs_N} and \eqref{Lambda} in \eqref{HDE_UVcorection}, the energy density $\rho_H$ is expressed in terms of the number of e-fold, given by
\begin{equation}
\rho_H = 3 M_p^2 \; \Lambda^2_{UV} \; \left[ d_0 \left( {e^{-\beta N} \over 1 + {\beta \over 3}} + 1 \right) \right]^{2/3}.
\end{equation}
It is illustrated in Fig.\ref{energy_scale_inflation} versus the number of e-fold $N$ for different values of the constant $\beta$. The energy scale of inflation is of the order of $10^{16} \; {\rm GeV}$, and it decreases
as $\beta$ increases.
%%%%%%%%%%%%%%%%%%%%%%%%%%%%
%%%%%%%%%%%%%%%%%%%%%%%%%%%%
Table.\ref{table01} presents some numerical results of the model about the observable parameters $n_s$, $r$, and the energy scale of inflation. It is seen that as $\beta$ increase both the scalar spectral index and the tensor-to-scalar ratio decreases. Moreover, there is a small reduction in the energy scale of inflation, however, it does not change the order of the energy scale.
%%%%%%%%%%%%%%%%%%%%%%%%%%%%%%%%%%%%%%%%%%%%%%%%%%%
%\begin{widetext}

\begin{table}
\caption{\label{table01} The table shows the estimated values of $n_s$, $r$, and energy scale of inflation, i.e. $\rho_H^{1/4}$ for different values of $\beta$ and $d_0$, where $N=65$.}
\begin{tabular}{p{1cm}p{1.3cm}p{1.3cm}p{2.2cm}p{1.5cm}p{1.5cm}p{1.7cm}}
\hline
$N$ & $\ \ \beta$ & $\quad d_0$ & $\ \quad \Lambda_{UV}$ & $\quad n_s$ & $\quad \ r$  & energy scale \\[1mm]
\hline
$60$ & $0.013$ & $-10$     & $6.03 \times 10^{13}$  &  $0.9686$  &  $0.0142$  &  $1.98 \times 10^{16}$ \\[2mm]

$60$ & $0.013$ & $-100$   & $2.80 \times 10^{13}$  &  $0.9686$  &  $0.0142$  &  $1.98 \times 10^{16}$ \\[2mm]

$60$ & $0.013$ & $-1000$  & $1.30 \times 10^{13}$  &  $0.9686$  &  $0.0142$  & $ 1.98 \times 10^{16}$ \\[2mm]

$60$ & $0.023$ & $-10$    & $4.54 \times 10^{13}$  &  $0.9640$  &  $0.0099$  &  $1.81 \times 10^{16}$ \\[2mm]

$60$ & $0.023$ & $-100$   & $2.11 \times 10^{13}$  &  $0.9640$  &  $0.0099$  &  $1.81 \times 10^{16}$ \\[2mm]

$60$ & $0.023$ & $-1000$ & $9.79 \times 10^{12}$  &  $0.9640$  &  $0.0099$  &  $1.81 \times 10^{16}$ \\[2mm]
%+++++++++++++++++++++++++++++++++++++
$65$ & $0.018$ & $-10$     & $4.77 \times 10^{13}$  &  $0.9684$  &  $0.0104$  &  $1.83 \times 10^{16}$ \\[2mm]

$65$ & $0.018$ & $-100$  & $2.21 \times 10^{13}$  &  $0.9684$  &  $0.0104$  &  $1.83 \times 10^{16}$ \\[2mm]

$65$ & $0.018$ & $-1000$  & $1.02 \times 10^{13}$  &  $0.9684$  & $0.0104$  &  $1.83 \times 10^{16}$ \\[2mm]

$65$ & $0.028$ & $-10$   & $3.66 \times 10^{13}$  &  $0.9629$  &  $0.0069$  &  $1.66 \times 10^{16}$ \\[2mm]

$65$ & $0.028$ & $-100$ & $1.70 \times 10^{13}$  &  $0.9629$  &  $0.0069$  &  $1.66 \times 10^{16}$ \\[2mm]

$65$ & $0.028$ & $-1000$ & $7.89 \times 10^{12}$  &  $0.9629$  &  $0.0069$  &  $1.66 \times 10^{16}$ \\[2mm]

\hline
\end{tabular}
\end{table}

%\end{widetext}
%%%%%%%%%%%%%%%%%%%%%%%%%%%%%%%%%%%%%%%%%%%%%%%%%%%

%%%%%%%%%%%%%%%%%%%%%%%%%%%%%%%%%%%%%%%%%%
%%%%%%%%%%%%%%%%%%%%%%%%%%%%%%%%%%%%%%%%%%
%%%%%%%%%%%%%%%%%%%%%%%%%%%%%%%%%%%%%%%%%%
%%%%%%%%%%%%%%%%%%%%%%%%%%%%%%%%%%%%%%%%%%\\
%%%%%%%%%%%%%%%%%%%%%%%%%%%%%%%%%%%%%%%%%%
%%%%%%%%%%%%%%%%%%%%%%%%%%%%%%%%%%%%%%%%%%
%%%%%%%%%%%%%%%%%%%%%%%%%%%%%%%%%%%%%%%%%%
%%%%%%%%%%%%%%%%%%%%%%%%%%%%%%%%%%%%%%%%%%
%%%%%%%%%%%%%%%%%%%%%%%%%%%%%%%%%%%%%%%%%%
\section{Correspondence between HDE and scalar field}\label{HDE_scalarfield}
\begin{figure}[h]
\centering
\includegraphics[width=8cm]{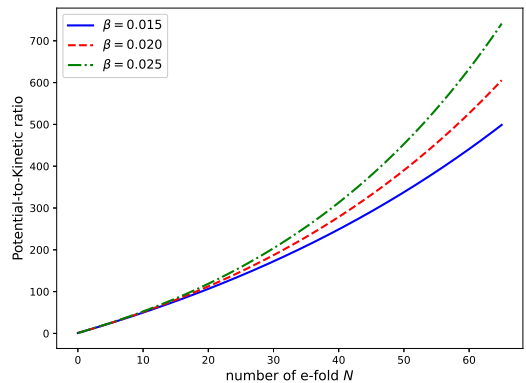}
\caption{\label{potential_to_kinetic} The figure illustrate the ratio of the potential-to-kinetic of the scalar field versus the number of e-fold for different values of $\beta$.}
\end{figure}

\begin{figure}[t]
\centering
\includegraphics[width=8cm]{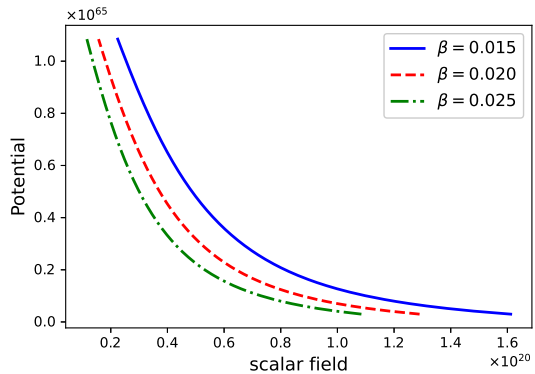}
\caption{\label{potential_scalarfield} The potential versus the scalar field is plotted during inflation for different values of $\beta$; and $d_0 = -10$.}
\end{figure}

The HDE has been widely used as a candidate of dark energy in describing the universe evolution in late times. It has been a common treat to correspond the HDE to a scalar field and reconstruct a potential
\cite{Chattopadhyay:2014yda,Chattopadhyay:2016enn,Samanta:2013vna,Sheykhi:2011,Yang:2011zza,
Karami:2009we}. In this regard, the energy density of the canonical scalar field is taken equal to the energy density of the HDE, i.e.
\begin{equation}
\rho_\phi = {1 \over 2} \; \dot\phi^2 + V(\phi) = \rho_{HDE} .
\end{equation}
Then, the potential is read as
\begin{equation}\label{pot_HDE}
V(\phi) = \rho_{HDE} - {1 \over 2} \; \dot\phi^2.
\end{equation}
On the other hand, from the second Friedmann equation,
\begin{equation}
-2 M_p^2 \; \dot{H} = \rho + p ,
\end{equation}
for the scalar field and HDE, respectively, one has
\begin{eqnarray}
-2 M_p^2 \; \dot{H} & = & \dot{\phi}^2 , \nonumber \\
-2 M_p^2 \; \dot{H} & = & \rho_{HDE} + p_{HDE} = -2 M_p^2 \; \Lambda_{UV}  \; {c \; \dot{c} \over \sqrt{c^2 - 1}}. \qquad
\end{eqnarray}
where Eq.\eqref{HDE_Friedmann} has been utilized for the second relation. Therefore, the kinetic term of the scalar field is read as
\begin{equation}\label{scalarfield_kinetic}
\dot{\phi}^2 = {-2 \over 3} M_p^2  \; \Lambda^2_{UV} \; \beta \;
              { \Big( (c^2 - 1)^{3/2} + d_0 \Big) \over \sqrt{c^2 - 1}},
\end{equation}
where Eq.\eqref{c_vs_N} has been used. Substituting above kinetic term in the relation of the potential, and by using Eq.\eqref{epsilon1_vs_c}, one arrives at
\begin{equation}
V(\phi) = 3 M_p^2 \; \Lambda^2_{UV} \; \big( c^2 - 1 \big) \;
    \left( \; 1 - {1 \over 3} \; \epsilon_1 \; \right).
\end{equation}
By applying Eq.\eqref{c_vs_N}, both the kinetic term and the potential are obtained in terms of the number of e-fold. Then, they could be estimated in any time of inflation. Fig.\ref{potential_to_kinetic} displays the potential-to-kinetic ratio versus the number of e-fold for different values of the constant $\beta$.
%%%%%%%%%%%%%%%%%%%%%%%%%%%%%%5
%%%%%%%%%%%%%%%%%%%%%%%%%%%%%%
It is realized that the early times of inflation, where number of e-fold is about $55-65$, the potential is much bigger than the kinetic term of the scalar field. This is what we expect in the slow-roll inflation. The ratio becomes bigger by choosing higher values of $\beta$. By approaching to the end of inflation, the ratio gets smaller and the kinetic term comes close to the potential. \\
Taking integrate from Eq.\eqref{scalarfield_kinetic} and by utilizing Eq.\eqref{c_vs_N}, the scalar field is obtained in terms of the number of e-fold. Then, by parametric plot, one could depict the potential versus the scalar field in the time of inflation, as presented in Fig.\ref{potential_scalarfield}.
%%%%%%%%%%%%%%%%%%%%%%%%%%%%%%5
%%%%%%%%%%%%%%%%%%%%%%%%%%%%%%
It is found that inflation begins at smaller values of the scalar field. The scalar field rolls down slowly from the top of the potential and grows in magnitude by approaching to the end of inflation.

%%%%%%%%%%%%%%%%%%%%%%%%%%%%%%%%%%%%%%%%%%
%%%%%%%%%%%%%%%%%%%%%%%%%%%%%%%%%%%%%%%%%%
%%%%%%%%%%%%%%%%%%%%%%%%%%%%%%%%%%%%%%%%%%
%%%%%%%%%%%%%%%%%%%%%%%%%%%%%%%%%%%%%%%%%%
%%%%%%%%%%%%%%%%%%%%%%%%%%%%%%%%%%%%%%%%%%
\section{Attractor behavior of the solution}\label{HDE_scalarfield}
The attractor behavior of the solution could be investigated both analytically and numerically. The analytical procedure of considering the attractor behavior is mostly used in the inflationary model with the Hamilton-Jacobi approach, where instead of the potential, the Hubble parameter is introduced as a function of the scalar field and the scalar field plays the role of time. In this method, a homogeneous perturbation $\delta H$ is assumed for the solution $H_0$, i.e. $H = H_0 + \delta H$. Substituting this in the Hamilton-Jacobi equation, one could obtain an evolution equation for the perturbation $\delta H$. If the perturbation $\delta H$ decays the solution $H_0$ is assumed to be stable. On the other hand, in the numerical procedure, the equation of motion of the scalar field is solved numerically where the Hubble parameter is inserted from the Friedmann equation. Then, the attractor behavior is considered by plotting the phase space diagram. \\
Here, by finding the parameter $c$, we actually obtain the Hubble parameter through Eq.(17). Then, From Eq.(32), we could actually have Hamilton Jacobi equation as 
\begin{equation}
V = 3M_p^2 \; H^2 + 2 M_p^2 \; \dot{H},
\end{equation}
where Eq.(34) has been used. Substituting $H = H_0 + \delta H$ in the above relation, one arrives at
\begin{equation}
\dot{\delta H} = - 3 H_0 \; \delta H ,
\end{equation} 
and the result will be 
\begin{equation}
\delta H = \delta H_i \; exp\left( -3 \int_{t_\star}^{t_e} H \; dt \right) = \delta H_i \; exp\left( -3 N \right).
\end{equation}
where $\delta H_i$ is the initial value of the perturbation $\delta H$. It is clear that by passing the time, the perturbation decays, therefore the solution will be stable.

%%%%%%%%%%%%%%%%%%%%%%%%%%%%%%%%%%%%%%%%%%
%%%%%%%%%%%%%%%%%%%%%%%%%%%%%%%%%%%%%%%%%%
%%%%%%%%%%%%%%%%%%%%%%%%%%%%%%%%%%%%%%%%%%
%%%%%%%%%%%%%%%%%%%%%%%%%%%%%%%%%%%%%%%%%%\\
%%%%%%%%%%%%%%%%%%%%%%%%%%%%%%%%%%%%%%%%%%
%%%%%%%%%%%%%%%%%%%%%%%%%%%%%%%%%%%%%%%%%%
%%%%%%%%%%%%%%%%%%%%%%%%%%%%%%%%%%%%%%%%%%
%%%%%%%%%%%%%%%%%%%%%%%%%%%%%%%%%%%%%%%%%%
%%%%%%%%%%%%%%%%%%%%%%%%%%%%%%%%%%%%%%%%%%
\section{Conclusion}\label{conclusion}
The original HDE, presented by $3c^2 M_p^2 / L^2$, with constant parameter $c$ and Hubble radius as the infrared cutoff, could not provide a suitable description for the late time acceleration of the universe. Then, in most holographic versions of inflation, different types of horizons are used as an infrared cutoff. However, the situation changes as one takes $c$ as a varying parameter. In this case, the original HDE with Hubble radius as infrared cutoff could present a suitable late time explanation. \\
There is a growing interest in the application of HDE in the early times of the universe. The recent studies indicate that HDE could have a suitable description for the slow-roll inflation. Then, it was found interesting to consider the scenario of constant-roll inflation originated by HDE with varying $c$ parameter and Hubble radius as the infrared cutoff. One of the features of constant-roll inflation is the constancy of the second slow-roll parameter. Then, applying the constancy of the second slow-roll parameter, a differential equation for the model was derived. An analytical solution was found, which was used to estimate the observable parameters at inflation. \\
The scalar spectral index and the tensor-to-scalar ratio were obtained to depend only on the constant $\beta$ and the number of e-fold. The $r-n_s$ resulted curves cross the interest area predicted by the data. Then, using Python coding and by applying the data about $n_s$ and $r$, we found a set of $(\beta, N)$ points which brings the model comparable to data. The set was plotted in Fig.\ref{beta_N_space} forming a parametric space. On the other hand, applying the result for $\beta$, and using the data for the amplitude of the scalar perturbations, we could also determine the constant $\Lambda_{UV}$. Gathering the result about the constants of the model, the energy scale of inflation was estimated to be of the order of $10^{16} \; {\rm GeV}$. \\
Finally, the HDE corresponded to a canonical scalar field. From, the reconstructed kinetic energy density and potential, it was found that the kinetic term during inflation is much smaller than the potential one, which is the thing we expect in slow-roll inflation.  The behavior of the potential versus the scalar field was depicted during inflation, which shows that inflation begins at small values of the scalar field. It slowly rolls down from the top of the potential. Then, inflation ends for larger values of the scalar field.

%\section*{Acknowledgments}
%The work of A. Mohammadi

\bibliography{CRI_HDE_Refs}

\begin{thebibliography}{94}%
\makeatletter
\providecommand \@ifxundefined [1]{%
 \@ifx{#1\undefined}
}%
\providecommand \@ifnum [1]{%
 \ifnum #1\expandafter \@firstoftwo
 \else \expandafter \@secondoftwo
 \fi
}%
\providecommand \@ifx [1]{%
 \ifx #1\expandafter \@firstoftwo
 \else \expandafter \@secondoftwo
 \fi
}%
\providecommand \natexlab [1]{#1}%
\providecommand \enquote  [1]{``#1''}%
\providecommand \bibnamefont  [1]{#1}%
\providecommand \bibfnamefont [1]{#1}%
\providecommand \citenamefont [1]{#1}%
\providecommand \href@noop [0]{\@secondoftwo}%
\providecommand \href [0]{\begingroup \@sanitize@url \@href}%
\providecommand \@href[1]{\@@startlink{#1}\@@href}%
\providecommand \@@href[1]{\endgroup#1\@@endlink}%
\providecommand \@sanitize@url [0]{\catcode `\\12\catcode `\$12\catcode
  `\&12\catcode `\#12\catcode `\^12\catcode `\_12\catcode `\%12\relax}%
\providecommand \@@startlink[1]{}%
\providecommand \@@endlink[0]{}%
\providecommand \url  [0]{\begingroup\@sanitize@url \@url }%
\providecommand \@url [1]{\endgroup\@href {#1}{\urlprefix }}%
\providecommand \urlprefix  [0]{URL }%
\providecommand \Eprint [0]{\href }%
\providecommand \doibase [0]{https://doi.org/}%
\providecommand \selectlanguage [0]{\@gobble}%
\providecommand \bibinfo  [0]{\@secondoftwo}%
\providecommand \bibfield  [0]{\@secondoftwo}%
\providecommand \translation [1]{[#1]}%
\providecommand \BibitemOpen [0]{}%
\providecommand \bibitemStop [0]{}%
\providecommand \bibitemNoStop [0]{.\EOS\space}%
\providecommand \EOS [0]{\spacefactor3000\relax}%
\providecommand \BibitemShut  [1]{\csname bibitem#1\endcsname}%
\let\auto@bib@innerbib\@empty
%</preamble>
\bibitem [{\citenamefont {Starobinsky}(1980)}]{starobinsky1980new}%
  \BibitemOpen
  \bibfield  {author} {\bibinfo {author} {\bibfnamefont {A.~A.}\ \bibnamefont
  {Starobinsky}},\ }\bibfield  {title} {\bibinfo {title} {A new type of
  isotropic cosmological models without singularity},\ }\href@noop {}
  {\bibfield  {journal} {\bibinfo  {journal} {Physics Letters B}\ }\textbf
  {\bibinfo {volume} {91}},\ \bibinfo {pages} {99} (\bibinfo {year}
  {1980})}\BibitemShut {NoStop}%
\bibitem [{\citenamefont {Guth}(1981)}]{Guth:1980zm}%
  \BibitemOpen
  \bibfield  {author} {\bibinfo {author} {\bibfnamefont {A.~H.}\ \bibnamefont
  {Guth}},\ }\bibfield  {title} {\bibinfo {title} {{The Inflationary Universe:
  A Possible Solution to the Horizon and Flatness Problems}},\ }\href
  {https://doi.org/10.1103/PhysRevD.23.347} {\bibfield  {journal} {\bibinfo
  {journal} {Phys. Rev.}\ }\textbf {\bibinfo {volume} {D23}},\ \bibinfo {pages}
  {347} (\bibinfo {year} {1981})},\ \bibinfo {note} {[Adv. Ser. Astrophys.
  Cosmol.3,139(1987)]}\BibitemShut {NoStop}%
%%CITATION = PHRVA,D23,347;%%
\bibitem [{\citenamefont {Akrami}\ \emph {et~al.}(2020)\citenamefont {Akrami}
  \emph {et~al.}}]{Planck:2018jri}%
  \BibitemOpen
  \bibfield  {author} {\bibinfo {author} {\bibfnamefont {Y.}~\bibnamefont
  {Akrami}} \emph {et~al.} (\bibinfo {collaboration} {Planck}),\ }\bibfield
  {title} {\bibinfo {title} {{Planck 2018 results. X. Constraints on
  inflation}},\ }\href {https://doi.org/10.1051/0004-6361/201833887} {\bibfield
   {journal} {\bibinfo  {journal} {Astron. Astrophys.}\ }\textbf {\bibinfo
  {volume} {641}},\ \bibinfo {pages} {A10} (\bibinfo {year} {2020})},\ \Eprint
  {https://arxiv.org/abs/1807.06211} {arXiv:1807.06211 [astro-ph.CO]}
  \BibitemShut {NoStop}%
\bibitem [{\citenamefont {Barenboim}\ and\ \citenamefont
  {Kinney}(2007)}]{Barenboim:2007ii}%
  \BibitemOpen
  \bibfield  {author} {\bibinfo {author} {\bibfnamefont {G.}~\bibnamefont
  {Barenboim}}\ and\ \bibinfo {author} {\bibfnamefont {W.~H.}\ \bibnamefont
  {Kinney}},\ }\bibfield  {title} {\bibinfo {title} {{Slow roll in simple
  non-canonical inflation}},\ }\href
  {https://doi.org/10.1088/1475-7516/2007/03/014} {\bibfield  {journal}
  {\bibinfo  {journal} {JCAP}\ }\textbf {\bibinfo {volume} {0703}},\ \bibinfo
  {pages} {014}},\ \Eprint {https://arxiv.org/abs/astro-ph/0701343}
  {arXiv:astro-ph/0701343 [astro-ph]} \BibitemShut {NoStop}%
%%CITATION = ASTRO-PH/0701343;%%
\bibitem [{\citenamefont {Franche}\ \emph {et~al.}(2010)\citenamefont
  {Franche}, \citenamefont {Gwyn}, \citenamefont {Underwood},\ and\
  \citenamefont {Wissanji}}]{Franche:2010yj}%
  \BibitemOpen
  \bibfield  {author} {\bibinfo {author} {\bibfnamefont {P.}~\bibnamefont
  {Franche}}, \bibinfo {author} {\bibfnamefont {R.}~\bibnamefont {Gwyn}},
  \bibinfo {author} {\bibfnamefont {B.}~\bibnamefont {Underwood}},\ and\
  \bibinfo {author} {\bibfnamefont {A.}~\bibnamefont {Wissanji}},\ }\bibfield
  {title} {\bibinfo {title} {{Initial Conditions for Non-Canonical
  Inflation}},\ }\href {https://doi.org/10.1103/PhysRevD.82.063528} {\bibfield
  {journal} {\bibinfo  {journal} {Phys. Rev.}\ }\textbf {\bibinfo {volume}
  {D82}},\ \bibinfo {pages} {063528} (\bibinfo {year} {2010})},\ \Eprint
  {https://arxiv.org/abs/1002.2639} {arXiv:1002.2639 [hep-th]} \BibitemShut
  {NoStop}%
%%CITATION = ARXIV:1002.2639;%%
\bibitem [{\citenamefont {Unnikrishnan}\ \emph {et~al.}(2012)\citenamefont
  {Unnikrishnan}, \citenamefont {Sahni},\ and\ \citenamefont
  {Toporensky}}]{Unnikrishnan:2012zu}%
  \BibitemOpen
  \bibfield  {author} {\bibinfo {author} {\bibfnamefont {S.}~\bibnamefont
  {Unnikrishnan}}, \bibinfo {author} {\bibfnamefont {V.}~\bibnamefont
  {Sahni}},\ and\ \bibinfo {author} {\bibfnamefont {A.}~\bibnamefont
  {Toporensky}},\ }\bibfield  {title} {\bibinfo {title} {{Refining inflation
  using non-canonical scalars}},\ }\href
  {https://doi.org/10.1088/1475-7516/2012/08/018} {\bibfield  {journal}
  {\bibinfo  {journal} {JCAP}\ }\textbf {\bibinfo {volume} {1208}},\ \bibinfo
  {pages} {018}},\ \Eprint {https://arxiv.org/abs/1205.0786} {arXiv:1205.0786
  [astro-ph.CO]} \BibitemShut {NoStop}%
%%CITATION = ARXIV:1205.0786;%%
\bibitem [{\citenamefont {Rezazadeh}\ \emph {et~al.}(2015)\citenamefont
  {Rezazadeh}, \citenamefont {Karami},\ and\ \citenamefont
  {Karimi}}]{Rezazadeh:2014fwa}%
  \BibitemOpen
  \bibfield  {author} {\bibinfo {author} {\bibfnamefont {K.}~\bibnamefont
  {Rezazadeh}}, \bibinfo {author} {\bibfnamefont {K.}~\bibnamefont {Karami}},\
  and\ \bibinfo {author} {\bibfnamefont {P.}~\bibnamefont {Karimi}},\
  }\bibfield  {title} {\bibinfo {title} {{Intermediate inflation from a
  non-canonical scalar field}},\ }\href
  {https://doi.org/$10.1088/1475-7516/2015/09/053$} {\bibfield  {journal}
  {\bibinfo  {journal} {JCAP}\ }\textbf {\bibinfo {volume} {1509}}\bibfield
  {number} {\bibinfo  {number} { (09)},\ \bibinfo {pages} {053}},\ }\Eprint
  {https://arxiv.org/abs/1411.7302} {arXiv:1411.7302 [gr-qc]} \BibitemShut
  {NoStop}%
%%CITATION = ARXIV:1411.7302;%%
\bibitem [{\citenamefont {Saaidi}\ \emph {et~al.}(2015)\citenamefont {Saaidi},
  \citenamefont {Mohammadi},\ and\ \citenamefont {Golanbari}}]{Saaidi:2015kaa}%
  \BibitemOpen
  \bibfield  {author} {\bibinfo {author} {\bibfnamefont {K.}~\bibnamefont
  {Saaidi}}, \bibinfo {author} {\bibfnamefont {A.}~\bibnamefont {Mohammadi}},\
  and\ \bibinfo {author} {\bibfnamefont {T.}~\bibnamefont {Golanbari}},\
  }\bibfield  {title} {\bibinfo {title} {{Light of Planck-2015 on Noncanonical
  Inflation}},\ }\href {https://doi.org/10.1155/2015/926807} {\bibfield
  {journal} {\bibinfo  {journal} {Adv. High Energy Phys.}\ }\textbf {\bibinfo
  {volume} {2015}},\ \bibinfo {pages} {926807} (\bibinfo {year} {2015})},\
  \Eprint {https://arxiv.org/abs/1708.03675} {arXiv:1708.03675 [gr-qc]}
  \BibitemShut {NoStop}%
%%CITATION = ARXIV:1708.03675;%%
\bibitem [{\citenamefont {Fairbairn}\ and\ \citenamefont
  {Tytgat}(2002)}]{Fairbairn:2002yp}%
  \BibitemOpen
  \bibfield  {author} {\bibinfo {author} {\bibfnamefont {M.}~\bibnamefont
  {Fairbairn}}\ and\ \bibinfo {author} {\bibfnamefont {M.~H.~G.}\ \bibnamefont
  {Tytgat}},\ }\bibfield  {title} {\bibinfo {title} {{Inflation from a tachyon
  fluid?}},\ }\href {https://doi.org/10.1016/S0370-2693(02)02638-2} {\bibfield
  {journal} {\bibinfo  {journal} {Phys. Lett.}\ }\textbf {\bibinfo {volume}
  {B546}},\ \bibinfo {pages} {1} (\bibinfo {year} {2002})},\ \Eprint
  {https://arxiv.org/abs/hep-th/0204070} {arXiv:hep-th/0204070 [hep-th]}
  \BibitemShut {NoStop}%
%%CITATION = HEP-TH/0204070;%%
\bibitem [{\citenamefont {Mukohyama}(2002)}]{Mukohyama:2002cn}%
  \BibitemOpen
  \bibfield  {author} {\bibinfo {author} {\bibfnamefont {S.}~\bibnamefont
  {Mukohyama}},\ }\bibfield  {title} {\bibinfo {title} {{Brane cosmology driven
  by the rolling tachyon}},\ }\href
  {https://doi.org/$10.1103/PhysRevD.66.024009$} {\bibfield  {journal}
  {\bibinfo  {journal} {Phys. Rev.}\ }\textbf {\bibinfo {volume} {D66}},\
  \bibinfo {pages} {024009} (\bibinfo {year} {2002})},\ \Eprint
  {https://arxiv.org/abs/hep-th/0204084} {arXiv:hep-th/0204084 [hep-th]}
  \BibitemShut {NoStop}%
%%CITATION = HEP-TH/0204084;%%
\bibitem [{\citenamefont {Feinstein}(2002)}]{Feinstein:2002aj}%
  \BibitemOpen
  \bibfield  {author} {\bibinfo {author} {\bibfnamefont {A.}~\bibnamefont
  {Feinstein}},\ }\bibfield  {title} {\bibinfo {title} {{Power law inflation
  from the rolling tachyon}},\ }\href
  {https://doi.org/10.1103/PhysRevD.66.063511} {\bibfield  {journal} {\bibinfo
  {journal} {Phys. Rev.}\ }\textbf {\bibinfo {volume} {D66}},\ \bibinfo {pages}
  {063511} (\bibinfo {year} {2002})},\ \Eprint
  {https://arxiv.org/abs/hep-th/0204140} {arXiv:hep-th/0204140 [hep-th]}
  \BibitemShut {NoStop}%
%%CITATION = HEP-TH/0204140;%%
\bibitem [{\citenamefont {Padmanabhan}(2002)}]{Padmanabhan:2002cp}%
  \BibitemOpen
  \bibfield  {author} {\bibinfo {author} {\bibfnamefont {T.}~\bibnamefont
  {Padmanabhan}},\ }\bibfield  {title} {\bibinfo {title} {{Accelerated
  expansion of the universe driven by tachyonic matter}},\ }\href
  {https://doi.org/10.1103/PhysRevD.66.021301} {\bibfield  {journal} {\bibinfo
  {journal} {Phys. Rev.}\ }\textbf {\bibinfo {volume} {D66}},\ \bibinfo {pages}
  {021301} (\bibinfo {year} {2002})},\ \Eprint
  {https://arxiv.org/abs/hep-th/0204150} {arXiv:hep-th/0204150 [hep-th]}
  \BibitemShut {NoStop}%
%%CITATION = HEP-TH/0204150;%%
\bibitem [{\citenamefont {Aghamohammadi}\ \emph {et~al.}(2014)\citenamefont
  {Aghamohammadi}, \citenamefont {Mohammadi}, \citenamefont {Golanbari},\ and\
  \citenamefont {Saaidi}}]{Aghamohammadi:2014aca}%
  \BibitemOpen
  \bibfield  {author} {\bibinfo {author} {\bibfnamefont {A.}~\bibnamefont
  {Aghamohammadi}}, \bibinfo {author} {\bibfnamefont {A.}~\bibnamefont
  {Mohammadi}}, \bibinfo {author} {\bibfnamefont {T.}~\bibnamefont
  {Golanbari}},\ and\ \bibinfo {author} {\bibfnamefont {K.}~\bibnamefont
  {Saaidi}},\ }\bibfield  {title} {\bibinfo {title} {{Hamilton-Jacobi formalism
  for tachyon inflation}},\ }\href {https://doi.org/10.1103/PhysRevD.90.084028}
  {\bibfield  {journal} {\bibinfo  {journal} {Phys. Rev.}\ }\textbf {\bibinfo
  {volume} {D90}},\ \bibinfo {pages} {084028} (\bibinfo {year} {2014})},\
  \Eprint {https://arxiv.org/abs/1502.07578} {arXiv:1502.07578 [gr-qc]}
  \BibitemShut {NoStop}%
%%CITATION = ARXIV:1502.07578;%%
\bibitem [{\citenamefont {Spalinski}(2007)}]{Spalinski:2007dv}%
  \BibitemOpen
  \bibfield  {author} {\bibinfo {author} {\bibfnamefont {M.}~\bibnamefont
  {Spalinski}},\ }\bibfield  {title} {\bibinfo {title} {{On Power law inflation
  in DBI models}},\ }\href {https://doi.org/10.1088/1475-7516/2007/05/017}
  {\bibfield  {journal} {\bibinfo  {journal} {JCAP}\ }\textbf {\bibinfo
  {volume} {0705}},\ \bibinfo {pages} {017}},\ \Eprint
  {https://arxiv.org/abs/hep-th/0702196} {arXiv:hep-th/0702196 [hep-th]}
  \BibitemShut {NoStop}%
%%CITATION = HEP-TH/0702196;%%
\bibitem [{\citenamefont {Bessada}\ \emph {et~al.}(2009)\citenamefont
  {Bessada}, \citenamefont {Kinney},\ and\ \citenamefont
  {Tzirakis}}]{Bessada:2009pe}%
  \BibitemOpen
  \bibfield  {author} {\bibinfo {author} {\bibfnamefont {D.}~\bibnamefont
  {Bessada}}, \bibinfo {author} {\bibfnamefont {W.~H.}\ \bibnamefont
  {Kinney}},\ and\ \bibinfo {author} {\bibfnamefont {K.}~\bibnamefont
  {Tzirakis}},\ }\bibfield  {title} {\bibinfo {title} {{Inflationary potentials
  in DBI models}},\ }\href {https://doi.org/10.1088/1475-7516/2009/09/031}
  {\bibfield  {journal} {\bibinfo  {journal} {JCAP}\ }\textbf {\bibinfo
  {volume} {0909}},\ \bibinfo {pages} {031}},\ \Eprint
  {https://arxiv.org/abs/0907.1311} {arXiv:0907.1311 [gr-qc]} \BibitemShut
  {NoStop}%
%%CITATION = ARXIV:0907.1311;%%
\bibitem [{\citenamefont {Weller}\ \emph {et~al.}(2012)\citenamefont {Weller},
  \citenamefont {van~de Bruck},\ and\ \citenamefont {Mota}}]{Weller:2011ey}%
  \BibitemOpen
  \bibfield  {author} {\bibinfo {author} {\bibfnamefont {J.~M.}\ \bibnamefont
  {Weller}}, \bibinfo {author} {\bibfnamefont {C.}~\bibnamefont {van~de
  Bruck}},\ and\ \bibinfo {author} {\bibfnamefont {D.~F.}\ \bibnamefont
  {Mota}},\ }\bibfield  {title} {\bibinfo {title} {{Inflationary predictions in
  scalar-tensor DBI inflation}},\ }\href
  {https://doi.org/10.1088/1475-7516/2012/06/002} {\bibfield  {journal}
  {\bibinfo  {journal} {JCAP}\ }\textbf {\bibinfo {volume} {1206}},\ \bibinfo
  {pages} {002}},\ \Eprint {https://arxiv.org/abs/1111.0237} {arXiv:1111.0237
  [astro-ph.CO]} \BibitemShut {NoStop}%
%%CITATION = ARXIV:1111.0237;%%
\bibitem [{\citenamefont {Nazavari}\ \emph {et~al.}(2016)\citenamefont
  {Nazavari}, \citenamefont {Mohammadi}, \citenamefont {Ossoulian},\ and\
  \citenamefont {Saaidi}}]{Nazavari:2016yaa}%
  \BibitemOpen
  \bibfield  {author} {\bibinfo {author} {\bibfnamefont {N.}~\bibnamefont
  {Nazavari}}, \bibinfo {author} {\bibfnamefont {A.}~\bibnamefont {Mohammadi}},
  \bibinfo {author} {\bibfnamefont {Z.}~\bibnamefont {Ossoulian}},\ and\
  \bibinfo {author} {\bibfnamefont {K.}~\bibnamefont {Saaidi}},\ }\bibfield
  {title} {\bibinfo {title} {{Intermediate inflation driven by DBI scalar
  field}},\ }\href {https://doi.org/10.1103/PhysRevD.93.123504} {\bibfield
  {journal} {\bibinfo  {journal} {Phys. Rev.}\ }\textbf {\bibinfo {volume}
  {D93}},\ \bibinfo {pages} {123504} (\bibinfo {year} {2016})},\ \Eprint
  {https://arxiv.org/abs/1708.03676} {arXiv:1708.03676 [gr-qc]} \BibitemShut
  {NoStop}%
%%CITATION = ARXIV:1708.03676;%%
\bibitem [{\citenamefont {Maeda}\ and\ \citenamefont
  {Yamamoto}(2013)}]{maeda2013stability}%
  \BibitemOpen
  \bibfield  {author} {\bibinfo {author} {\bibfnamefont {K.-i.}\ \bibnamefont
  {Maeda}}\ and\ \bibinfo {author} {\bibfnamefont {K.}~\bibnamefont
  {Yamamoto}},\ }\bibfield  {title} {\bibinfo {title} {Stability analysis of
  inflation with an su (2) gauge field},\ }\href@noop {} {\bibfield  {journal}
  {\bibinfo  {journal} {Journal of Cosmology and Astroparticle Physics}\
  }\textbf {\bibinfo {volume} {2013}}\bibinfo  {number} { (12)},\ \bibinfo
  {pages} {018}}\BibitemShut {NoStop}%
\bibitem [{\citenamefont {Abolhasani}\ \emph {et~al.}(2014)\citenamefont
  {Abolhasani}, \citenamefont {Emami},\ and\ \citenamefont
  {Firouzjahi}}]{abolhasani2014primordial}%
  \BibitemOpen
\bibfield  {number} {  }\bibfield  {author} {\bibinfo {author} {\bibfnamefont
  {A.~A.}\ \bibnamefont {Abolhasani}}, \bibinfo {author} {\bibfnamefont
  {R.}~\bibnamefont {Emami}},\ and\ \bibinfo {author} {\bibfnamefont
  {H.}~\bibnamefont {Firouzjahi}},\ }\bibfield  {title} {\bibinfo {title}
  {Primordial anisotropies in gauged hybrid inflation},\ }\href@noop {}
  {\bibfield  {journal} {\bibinfo  {journal} {Journal of Cosmology and
  Astroparticle Physics}\ }\textbf {\bibinfo {volume} {2014}}\bibinfo  {number}
  { (05)},\ \bibinfo {pages} {016}}\BibitemShut {NoStop}%
\bibitem [{\citenamefont {Alexander}\ \emph {et~al.}(2015)\citenamefont
  {Alexander}, \citenamefont {Jyoti}, \citenamefont {Kosowsky},\ and\
  \citenamefont {Marcian{\`o}}}]{alexander2015dynamics}%
  \BibitemOpen
\bibfield  {number} {  }\bibfield  {author} {\bibinfo {author} {\bibfnamefont
  {S.}~\bibnamefont {Alexander}}, \bibinfo {author} {\bibfnamefont
  {D.}~\bibnamefont {Jyoti}}, \bibinfo {author} {\bibfnamefont
  {A.}~\bibnamefont {Kosowsky}},\ and\ \bibinfo {author} {\bibfnamefont
  {A.}~\bibnamefont {Marcian{\`o}}},\ }\bibfield  {title} {\bibinfo {title}
  {Dynamics of gauge field inflation},\ }\href@noop {} {\bibfield  {journal}
  {\bibinfo  {journal} {Journal of Cosmology and Astroparticle Physics}\
  }\textbf {\bibinfo {volume} {2015}}\bibinfo  {number} { (05)},\ \bibinfo
  {pages} {005}}\BibitemShut {NoStop}%
\bibitem [{\citenamefont {Tirandari}\ and\ \citenamefont
  {Saaidi}(2017)}]{tirandari2017anisotropic}%
  \BibitemOpen
\bibfield  {number} {  }\bibfield  {author} {\bibinfo {author} {\bibfnamefont
  {M.}~\bibnamefont {Tirandari}}\ and\ \bibinfo {author} {\bibfnamefont
  {K.}~\bibnamefont {Saaidi}},\ }\bibfield  {title} {\bibinfo {title}
  {Anisotropic inflation in brans--dicke gravity},\ }\href@noop {} {\bibfield
  {journal} {\bibinfo  {journal} {Nuclear Physics B}\ }\textbf {\bibinfo
  {volume} {925}},\ \bibinfo {pages} {403} (\bibinfo {year}
  {2017})}\BibitemShut {NoStop}%
\bibitem [{\citenamefont {Maartens}\ \emph {et~al.}(2000)\citenamefont
  {Maartens}, \citenamefont {Wands}, \citenamefont {Bassett},\ and\
  \citenamefont {Heard}}]{maartens2000chaotic}%
  \BibitemOpen
  \bibfield  {author} {\bibinfo {author} {\bibfnamefont {R.}~\bibnamefont
  {Maartens}}, \bibinfo {author} {\bibfnamefont {D.}~\bibnamefont {Wands}},
  \bibinfo {author} {\bibfnamefont {B.~A.}\ \bibnamefont {Bassett}},\ and\
  \bibinfo {author} {\bibfnamefont {I.~P.}\ \bibnamefont {Heard}},\ }\bibfield
  {title} {\bibinfo {title} {Chaotic inflation on the brane},\ }\href@noop {}
  {\bibfield  {journal} {\bibinfo  {journal} {Physical Review D}\ }\textbf
  {\bibinfo {volume} {62}},\ \bibinfo {pages} {041301} (\bibinfo {year}
  {2000})}\BibitemShut {NoStop}%
\bibitem [{\citenamefont {Golanbari}\ \emph {et~al.}(2014)\citenamefont
  {Golanbari}, \citenamefont {Mohammadi},\ and\ \citenamefont
  {Saaidi}}]{golanbari2014brane}%
  \BibitemOpen
  \bibfield  {author} {\bibinfo {author} {\bibfnamefont {T.}~\bibnamefont
  {Golanbari}}, \bibinfo {author} {\bibfnamefont {A.}~\bibnamefont
  {Mohammadi}},\ and\ \bibinfo {author} {\bibfnamefont {K.}~\bibnamefont
  {Saaidi}},\ }\bibfield  {title} {\bibinfo {title} {Brane inflation driven by
  noncanonical scalar field},\ }\href@noop {} {\bibfield  {journal} {\bibinfo
  {journal} {Physical Review D}\ }\textbf {\bibinfo {volume} {89}},\ \bibinfo
  {pages} {103529} (\bibinfo {year} {2014})}\BibitemShut {NoStop}%
\bibitem [{\citenamefont {Mohammadi}\ \emph
  {et~al.}(2020{\natexlab{a}})\citenamefont {Mohammadi}, \citenamefont
  {Golanbari}, \citenamefont {Nasri},\ and\ \citenamefont
  {Saaidi}}]{Mohammadi:2020ake}%
  \BibitemOpen
  \bibfield  {author} {\bibinfo {author} {\bibfnamefont {A.}~\bibnamefont
  {Mohammadi}}, \bibinfo {author} {\bibfnamefont {T.}~\bibnamefont
  {Golanbari}}, \bibinfo {author} {\bibfnamefont {S.}~\bibnamefont {Nasri}},\
  and\ \bibinfo {author} {\bibfnamefont {K.}~\bibnamefont {Saaidi}},\
  }\bibfield  {title} {\bibinfo {title} {{Brane inflation: Swampland Criteria,
  TCC, and Reheating constraint}},\ }\href@noop {} {\  (\bibinfo {year}
  {2020}{\natexlab{a}})},\ \Eprint {https://arxiv.org/abs/2006.09489}
  {arXiv:2006.09489 [gr-qc]} \BibitemShut {NoStop}%
\bibitem [{\citenamefont {Mohammadi}\ \emph
  {et~al.}(2020{\natexlab{b}})\citenamefont {Mohammadi}, \citenamefont
  {Golanbari},\ and\ \citenamefont {Enayati}}]{Mohammadi:2020ctd}%
  \BibitemOpen
  \bibfield  {author} {\bibinfo {author} {\bibfnamefont {A.}~\bibnamefont
  {Mohammadi}}, \bibinfo {author} {\bibfnamefont {T.}~\bibnamefont
  {Golanbari}},\ and\ \bibinfo {author} {\bibfnamefont {J.}~\bibnamefont
  {Enayati}},\ }\bibfield  {title} {\bibinfo {title} {{Brane inflation and
  Trans-Planckian censorship conjecture}},\ }\href@noop {} {\  (\bibinfo {year}
  {2020}{\natexlab{b}})},\ \Eprint {https://arxiv.org/abs/2012.01512}
  {arXiv:2012.01512 [hep-th]} \BibitemShut {NoStop}%
\bibitem [{\citenamefont {Berera}(1995)}]{berera1995warm}%
  \BibitemOpen
  \bibfield  {author} {\bibinfo {author} {\bibfnamefont {A.}~\bibnamefont
  {Berera}},\ }\bibfield  {title} {\bibinfo {title} {Warm inflation},\
  }\href@noop {} {\bibfield  {journal} {\bibinfo  {journal} {Physical Review
  Letters}\ }\textbf {\bibinfo {volume} {75}},\ \bibinfo {pages} {3218}
  (\bibinfo {year} {1995})}\BibitemShut {NoStop}%
\bibitem [{\citenamefont {Berera}(2000)}]{berera2000warm}%
  \BibitemOpen
  \bibfield  {author} {\bibinfo {author} {\bibfnamefont {A.}~\bibnamefont
  {Berera}},\ }\bibfield  {title} {\bibinfo {title} {Warm inflation in the
  adiabatic regime—a model, an existence proof for inflationary dynamics in
  quantum field theory},\ }\href@noop {} {\bibfield  {journal} {\bibinfo
  {journal} {Nuclear Physics B}\ }\textbf {\bibinfo {volume} {585}},\ \bibinfo
  {pages} {666} (\bibinfo {year} {2000})}\BibitemShut {NoStop}%
\bibitem [{\citenamefont {Hall}\ \emph {et~al.}(2004)\citenamefont {Hall},
  \citenamefont {Moss},\ and\ \citenamefont {Berera}}]{hall2004scalar}%
  \BibitemOpen
  \bibfield  {author} {\bibinfo {author} {\bibfnamefont {L.~M.}\ \bibnamefont
  {Hall}}, \bibinfo {author} {\bibfnamefont {I.~G.}\ \bibnamefont {Moss}},\
  and\ \bibinfo {author} {\bibfnamefont {A.}~\bibnamefont {Berera}},\
  }\bibfield  {title} {\bibinfo {title} {Scalar perturbation spectra from warm
  inflation},\ }\href@noop {} {\bibfield  {journal} {\bibinfo  {journal}
  {Physical Review D}\ }\textbf {\bibinfo {volume} {69}},\ \bibinfo {pages}
  {083525} (\bibinfo {year} {2004})}\BibitemShut {NoStop}%
\bibitem [{\citenamefont {Sayar}\ \emph {et~al.}(2017)\citenamefont {Sayar},
  \citenamefont {Mohammadi}, \citenamefont {Akhtari},\ and\ \citenamefont
  {Saaidi}}]{Sayar:2017pam}%
  \BibitemOpen
  \bibfield  {author} {\bibinfo {author} {\bibfnamefont {K.}~\bibnamefont
  {Sayar}}, \bibinfo {author} {\bibfnamefont {A.}~\bibnamefont {Mohammadi}},
  \bibinfo {author} {\bibfnamefont {L.}~\bibnamefont {Akhtari}},\ and\ \bibinfo
  {author} {\bibfnamefont {K.}~\bibnamefont {Saaidi}},\ }\bibfield  {title}
  {\bibinfo {title} {{Hamilton-Jacobi formalism to warm inflationary
  scenario}},\ }\href {https://doi.org/10.1103/PhysRevD.95.023501} {\bibfield
  {journal} {\bibinfo  {journal} {Phys. Rev.}\ }\textbf {\bibinfo {volume}
  {D95}},\ \bibinfo {pages} {023501} (\bibinfo {year} {2017})},\ \Eprint
  {https://arxiv.org/abs/1708.01714} {arXiv:1708.01714 [gr-qc]} \BibitemShut
  {NoStop}%
%%CITATION = ARXIV:1708.01714;%%
\bibitem [{\citenamefont {Akhtari}\ \emph {et~al.}(2017)\citenamefont
  {Akhtari}, \citenamefont {Mohammadi}, \citenamefont {Sayar},\ and\
  \citenamefont {Saaidi}}]{Akhtari:2017mxc}%
  \BibitemOpen
  \bibfield  {author} {\bibinfo {author} {\bibfnamefont {L.}~\bibnamefont
  {Akhtari}}, \bibinfo {author} {\bibfnamefont {A.}~\bibnamefont {Mohammadi}},
  \bibinfo {author} {\bibfnamefont {K.}~\bibnamefont {Sayar}},\ and\ \bibinfo
  {author} {\bibfnamefont {K.}~\bibnamefont {Saaidi}},\ }\bibfield  {title}
  {\bibinfo {title} {{Viscous warm inflation: Hamilton–Jacobi formalism}},\
  }\href {https://doi.org/10.1016/j.astropartphys.2017.02.002} {\bibfield
  {journal} {\bibinfo  {journal} {Astropart. Phys.}\ }\textbf {\bibinfo
  {volume} {90}},\ \bibinfo {pages} {28} (\bibinfo {year} {2017})},\ \Eprint
  {https://arxiv.org/abs/1710.05793} {arXiv:1710.05793 [astro-ph.CO]}
  \BibitemShut {NoStop}%
%%CITATION = ARXIV:1710.05793;%%
\bibitem [{\citenamefont {Sheikhahmadi}\ \emph {et~al.}(2019)\citenamefont
  {Sheikhahmadi}, \citenamefont {Mohammadi}, \citenamefont {Aghamohammadi},
  \citenamefont {Harko}, \citenamefont {Herrera}, \citenamefont {Corda},
  \citenamefont {Abebe},\ and\ \citenamefont {Saaidi}}]{Sheikhahmadi:2019gzs}%
  \BibitemOpen
  \bibfield  {author} {\bibinfo {author} {\bibfnamefont {H.}~\bibnamefont
  {Sheikhahmadi}}, \bibinfo {author} {\bibfnamefont {A.}~\bibnamefont
  {Mohammadi}}, \bibinfo {author} {\bibfnamefont {A.}~\bibnamefont
  {Aghamohammadi}}, \bibinfo {author} {\bibfnamefont {T.}~\bibnamefont
  {Harko}}, \bibinfo {author} {\bibfnamefont {R.}~\bibnamefont {Herrera}},
  \bibinfo {author} {\bibfnamefont {C.}~\bibnamefont {Corda}}, \bibinfo
  {author} {\bibfnamefont {A.}~\bibnamefont {Abebe}},\ and\ \bibinfo {author}
  {\bibfnamefont {K.}~\bibnamefont {Saaidi}},\ }\bibfield  {title} {\bibinfo
  {title} {{Constraining chameleon field driven warm inflation with Planck 2018
  data}},\ }\href {https://doi.org/10.1140/epjc/s10052-019-7571-0} {\bibfield
  {journal} {\bibinfo  {journal} {Eur. Phys. J.}\ }\textbf {\bibinfo {volume}
  {C79}},\ \bibinfo {pages} {1038} (\bibinfo {year} {2019})},\ \Eprint
  {https://arxiv.org/abs/1907.10966} {arXiv:1907.10966 [gr-qc]} \BibitemShut
  {NoStop}%
%%CITATION = ARXIV:1907.10966;%%
\bibitem [{\citenamefont {Mohammadi}\ \emph
  {et~al.}(2020{\natexlab{c}})\citenamefont {Mohammadi}, \citenamefont
  {Golanbari}, \citenamefont {Sheikhahmadi}, \citenamefont {Sayar},
  \citenamefont {Akhtari}, \citenamefont {Rasheed},\ and\ \citenamefont
  {Saaidi}}]{Rasheed:2020syk}%
  \BibitemOpen
  \bibfield  {author} {\bibinfo {author} {\bibfnamefont {A.}~\bibnamefont
  {Mohammadi}}, \bibinfo {author} {\bibfnamefont {T.}~\bibnamefont
  {Golanbari}}, \bibinfo {author} {\bibfnamefont {H.}~\bibnamefont
  {Sheikhahmadi}}, \bibinfo {author} {\bibfnamefont {K.}~\bibnamefont {Sayar}},
  \bibinfo {author} {\bibfnamefont {L.}~\bibnamefont {Akhtari}}, \bibinfo
  {author} {\bibfnamefont {M.}~\bibnamefont {Rasheed}},\ and\ \bibinfo {author}
  {\bibfnamefont {K.}~\bibnamefont {Saaidi}},\ }\bibfield  {title} {\bibinfo
  {title} {{Warm tachyon inflation and swampland criteria}},\ }\href
  {https://doi.org/10.1088/1674-1137/44/9/095101} {\bibfield  {journal}
  {\bibinfo  {journal} {Chin. Phys. C}\ }\textbf {\bibinfo {volume} {44}},\
  \bibinfo {pages} {095101} (\bibinfo {year} {2020}{\natexlab{c}})},\ \Eprint
  {https://arxiv.org/abs/2001.10042} {arXiv:2001.10042 [gr-qc]} \BibitemShut
  {NoStop}%
\bibitem [{\citenamefont {Mohammadi}\ \emph {et~al.}(2018)\citenamefont
  {Mohammadi}, \citenamefont {Saaidi},\ and\ \citenamefont
  {Golanbari}}]{Mohammadi:2018oku}%
  \BibitemOpen
  \bibfield  {author} {\bibinfo {author} {\bibfnamefont {A.}~\bibnamefont
  {Mohammadi}}, \bibinfo {author} {\bibfnamefont {K.}~\bibnamefont {Saaidi}},\
  and\ \bibinfo {author} {\bibfnamefont {T.}~\bibnamefont {Golanbari}},\
  }\bibfield  {title} {\bibinfo {title} {{Tachyon constant-roll inflation}},\
  }\href {https://doi.org/10.1103/PhysRevD.97.083006} {\bibfield  {journal}
  {\bibinfo  {journal} {Phys. Rev.}\ }\textbf {\bibinfo {volume} {D97}},\
  \bibinfo {pages} {083006} (\bibinfo {year} {2018})},\ \Eprint
  {https://arxiv.org/abs/1801.03487} {arXiv:1801.03487 [hep-ph]} \BibitemShut
  {NoStop}%
%%CITATION = ARXIV:1801.03487;%%
\bibitem [{\citenamefont {Mohammadi}\ \emph {et~al.}(2019)\citenamefont
  {Mohammadi}, \citenamefont {Saaidi},\ and\ \citenamefont
  {Sheikhahmadi}}]{Mohammadi:2019dpu}%
  \BibitemOpen
  \bibfield  {author} {\bibinfo {author} {\bibfnamefont {A.}~\bibnamefont
  {Mohammadi}}, \bibinfo {author} {\bibfnamefont {K.}~\bibnamefont {Saaidi}},\
  and\ \bibinfo {author} {\bibfnamefont {H.}~\bibnamefont {Sheikhahmadi}},\
  }\bibfield  {title} {\bibinfo {title} {{Constant-roll approach to
  non-canonical inflation}},\ }\href
  {https://doi.org/10.1103/PhysRevD.100.083520} {\bibfield  {journal} {\bibinfo
   {journal} {Phys. Rev.}\ }\textbf {\bibinfo {volume} {D100}},\ \bibinfo
  {pages} {083520} (\bibinfo {year} {2019})},\ \Eprint
  {https://arxiv.org/abs/1803.01715} {arXiv:1803.01715 [astro-ph.CO]}
  \BibitemShut {NoStop}%
%%CITATION = ARXIV:1803.01715;%%
\bibitem [{\citenamefont {Golanbari}\ \emph {et~al.}(2020)\citenamefont
  {Golanbari}, \citenamefont {Mohammadi},\ and\ \citenamefont
  {Saaidi}}]{Mohammadi:2018zkf}%
  \BibitemOpen
  \bibfield  {author} {\bibinfo {author} {\bibfnamefont {T.}~\bibnamefont
  {Golanbari}}, \bibinfo {author} {\bibfnamefont {A.}~\bibnamefont
  {Mohammadi}},\ and\ \bibinfo {author} {\bibfnamefont {K.}~\bibnamefont
  {Saaidi}},\ }\bibfield  {title} {\bibinfo {title} {{Observational constraints
  on DBI constant-roll inflation}},\ }\href
  {https://doi.org/10.1016/j.dark.2019.100456} {\bibfield  {journal} {\bibinfo
  {journal} {Phys. Dark Univ.}\ }\textbf {\bibinfo {volume} {27}},\ \bibinfo
  {pages} {100456} (\bibinfo {year} {2020})},\ \Eprint
  {https://arxiv.org/abs/1808.07246} {arXiv:1808.07246 [gr-qc]} \BibitemShut
  {NoStop}%
%%CITATION = ARXIV:1808.07246;%%
\bibitem [{\citenamefont {Mohammadi}\ \emph
  {et~al.}(2020{\natexlab{d}})\citenamefont {Mohammadi}, \citenamefont
  {Golanbari},\ and\ \citenamefont {Saaidi}}]{Mohammadi:2019qeu}%
  \BibitemOpen
  \bibfield  {author} {\bibinfo {author} {\bibfnamefont {A.}~\bibnamefont
  {Mohammadi}}, \bibinfo {author} {\bibfnamefont {T.}~\bibnamefont
  {Golanbari}},\ and\ \bibinfo {author} {\bibfnamefont {K.}~\bibnamefont
  {Saaidi}},\ }\bibfield  {title} {\bibinfo {title} {{Beta-function formalism
  for k-essence constant-roll inflation}},\ }\href
  {https://doi.org/10.1016/j.dark.2020.100505} {\bibfield  {journal} {\bibinfo
  {journal} {Phys. Dark Univ.}\ }\textbf {\bibinfo {volume} {28}},\ \bibinfo
  {pages} {100505} (\bibinfo {year} {2020}{\natexlab{d}})},\ \Eprint
  {https://arxiv.org/abs/1912.07006} {arXiv:1912.07006 [gr-qc]} \BibitemShut
  {NoStop}%
%%CITATION = ARXIV:1912.07006;%%
\bibitem [{Moh()}]{Mohammadi:2020ftb}%
  \BibitemOpen
  \href@noop {} {\ }\BibitemShut {NoStop}%
\bibitem [{\citenamefont {Linde}(2000)}]{Linde:2000kn}%
  \BibitemOpen
  \bibfield  {author} {\bibinfo {author} {\bibfnamefont {A.~D.}\ \bibnamefont
  {Linde}},\ }\bibfield  {title} {\bibinfo {title} {{Inflationary cosmology}},\
  }\href {https://doi.org/10.1016/S0370-1573(00)00038-7} {\bibfield  {journal}
  {\bibinfo  {journal} {Phys. Rept.}\ }\textbf {\bibinfo {volume} {333}},\
  \bibinfo {pages} {575} (\bibinfo {year} {2000})}\BibitemShut {NoStop}%
%%CITATION = PRPLC,333,575;%%
\bibitem [{\citenamefont {Linde}(1990)}]{Linde:2005ht}%
  \BibitemOpen
  \bibfield  {author} {\bibinfo {author} {\bibfnamefont {A.~D.}\ \bibnamefont
  {Linde}},\ }\bibfield  {title} {\bibinfo {title} {{Particle physics and
  inflationary cosmology}},\ }\href@noop {} {\bibfield  {journal} {\bibinfo
  {journal} {Contemp. Concepts Phys.}\ }\textbf {\bibinfo {volume} {5}},\
  \bibinfo {pages} {1} (\bibinfo {year} {1990})},\ \Eprint
  {https://arxiv.org/abs/hep-th/0503203} {arXiv:hep-th/0503203 [hep-th]}
  \BibitemShut {NoStop}%
%%CITATION = HEP-TH/0503203;%%
\bibitem [{\citenamefont {Linde}(2005{\natexlab{a}})}]{Linde:2005vy}%
  \BibitemOpen
  \bibfield  {author} {\bibinfo {author} {\bibfnamefont {A.~D.}\ \bibnamefont
  {Linde}},\ }\bibfield  {title} {\bibinfo {title} {{Current understanding of
  inflation}},\ }\bibfield  {booktitle} {\emph {\bibinfo {booktitle}
  {{Proceedings, 6th UCLA Symposium on Sources and Detection of Dark Matter and
  Dark Energy in the Universe: Marina del Rey, CA, USA, February 18-20,
  2004}}},\ }\href {https://doi.org/10.1016/j.newar.2005.01.002} {\bibfield
  {journal} {\bibinfo  {journal} {New Astron. Rev.}\ }\textbf {\bibinfo
  {volume} {49}},\ \bibinfo {pages} {35} (\bibinfo {year}
  {2005}{\natexlab{a}})}\BibitemShut {NoStop}%
%%CITATION = ASTRE,49,35;%%
\bibitem [{\citenamefont {Linde}(2005{\natexlab{b}})}]{Linde:2004kg}%
  \BibitemOpen
  \bibfield  {author} {\bibinfo {author} {\bibfnamefont {A.~D.}\ \bibnamefont
  {Linde}},\ }\bibfield  {title} {\bibinfo {title} {{Prospects of inflation}},\
  }\bibfield  {booktitle} {\emph {\bibinfo {booktitle} {{Proceedings, Nobel
  Symposium 127 on String theory and cosmology: Sigtuna, Sweden, August 14-19,
  2003}}},\ }\href {https://doi.org/10.1238/Physica.Topical.117a00040}
  {\bibfield  {journal} {\bibinfo  {journal} {Phys. Scripta}\ }\textbf
  {\bibinfo {volume} {T117}},\ \bibinfo {pages} {40} (\bibinfo {year}
  {2005}{\natexlab{b}})},\ \Eprint {https://arxiv.org/abs/hep-th/0402051}
  {arXiv:hep-th/0402051 [hep-th]} \BibitemShut {NoStop}%
%%CITATION = HEP-TH/0402051;%%
\bibitem [{\citenamefont {Riotto}(2003)}]{Riotto:2002yw}%
  \BibitemOpen
  \bibfield  {author} {\bibinfo {author} {\bibfnamefont {A.}~\bibnamefont
  {Riotto}},\ }\bibfield  {title} {\bibinfo {title} {{Inflation and the theory
  of cosmological perturbations}},\ }\bibfield  {booktitle} {\emph {\bibinfo
  {booktitle} {{Astroparticle physics and cosmology. Proceedings: Summer
  School, Trieste, Italy, Jun 17-Jul 5 2002}}},\ }\href@noop {} {\bibfield
  {journal} {\bibinfo  {journal} {ICTP Lect. Notes Ser.}\ }\textbf {\bibinfo
  {volume} {14}},\ \bibinfo {pages} {317} (\bibinfo {year} {2003})},\ \Eprint
  {https://arxiv.org/abs/hep-ph/0210162} {arXiv:hep-ph/0210162 [hep-ph]}
  \BibitemShut {NoStop}%
%%CITATION = HEP-PH/0210162;%%
\bibitem [{\citenamefont {Baumann}(2011)}]{Baumann:2009ds}%
  \BibitemOpen
  \bibfield  {author} {\bibinfo {author} {\bibfnamefont {D.}~\bibnamefont
  {Baumann}},\ }\bibfield  {title} {\bibinfo {title} {{TASI Lecture on
  Inflation}},\ }in\ \href {https://doi.org/$10.1142/9789814327183_0010$}
  {\emph {\bibinfo {booktitle} {{Physics of the large and the small, TASI 09,
  proceedings of the Theoretical Advanced Study Institute in Elementary
  Particle Physics, Boulder, Colorado, USA, 1-26 June 2009}}}}\ (\bibinfo
  {year} {2011})\ pp.\ \bibinfo {pages} {523--686},\ \Eprint
  {https://arxiv.org/abs/0907.5424} {arXiv:0907.5424 [hep-th]} \BibitemShut
  {NoStop}%
%%CITATION = ARXIV:0907.5424;%%
\bibitem [{\citenamefont {Weinberg}(2008)}]{Weinberg:2008zzc}%
  \BibitemOpen
  \bibfield  {author} {\bibinfo {author} {\bibfnamefont {S.}~\bibnamefont
  {Weinberg}},\ }\href {http://www.oup.com/uk/catalogue/?ci=9780198526827}
  {\emph {\bibinfo {title} {{Cosmology}}}}\ (\bibinfo {year}
  {2008})\BibitemShut {NoStop}%
%%CITATION = INSPIRE-794379;%%
\bibitem [{\citenamefont {Lyth}\ and\ \citenamefont
  {Liddle}(2009)}]{Lyth:2009zz}%
  \BibitemOpen
  \bibfield  {author} {\bibinfo {author} {\bibfnamefont {D.~H.}\ \bibnamefont
  {Lyth}}\ and\ \bibinfo {author} {\bibfnamefont {A.~R.}\ \bibnamefont
  {Liddle}},\ }\href
  {http://www.cambridge.org/uk/catalogue/catalogue.asp?isbn=9780521828499}
  {\emph {\bibinfo {title} {{The primordial density perturbation: Cosmology,
  inflation and the origin of structure}}}}\ (\bibinfo {year}
  {2009})\BibitemShut {NoStop}%
%%CITATION = INSPIRE-853992;%%
\bibitem [{\citenamefont {Kinney}(2005)}]{Kinney:2005vj}%
  \BibitemOpen
  \bibfield  {author} {\bibinfo {author} {\bibfnamefont {W.~H.}\ \bibnamefont
  {Kinney}},\ }\bibfield  {title} {\bibinfo {title} {{Horizon crossing and
  inflation with large eta}},\ }\href
  {https://doi.org/10.1103/PhysRevD.72.023515} {\bibfield  {journal} {\bibinfo
  {journal} {Phys. Rev.}\ }\textbf {\bibinfo {volume} {D72}},\ \bibinfo {pages}
  {023515} (\bibinfo {year} {2005})},\ \Eprint
  {https://arxiv.org/abs/gr-qc/0503017} {arXiv:gr-qc/0503017 [gr-qc]}
  \BibitemShut {NoStop}%
%%CITATION = GR-QC/0503017;%%
\bibitem [{\citenamefont {Martin}\ \emph {et~al.}(2013)\citenamefont {Martin},
  \citenamefont {Motohashi},\ and\ \citenamefont {Suyama}}]{Martin:2012pe}%
  \BibitemOpen
  \bibfield  {author} {\bibinfo {author} {\bibfnamefont {J.}~\bibnamefont
  {Martin}}, \bibinfo {author} {\bibfnamefont {H.}~\bibnamefont {Motohashi}},\
  and\ \bibinfo {author} {\bibfnamefont {T.}~\bibnamefont {Suyama}},\
  }\bibfield  {title} {\bibinfo {title} {{Ultra Slow-Roll Inflation and the
  non-Gaussianity Consistency Relation}},\ }\href
  {https://doi.org/10.1103/PhysRevD.87.023514} {\bibfield  {journal} {\bibinfo
  {journal} {Phys. Rev.}\ }\textbf {\bibinfo {volume} {D87}},\ \bibinfo {pages}
  {023514} (\bibinfo {year} {2013})},\ \Eprint
  {https://arxiv.org/abs/1211.0083} {arXiv:1211.0083 [astro-ph.CO]}
  \BibitemShut {NoStop}%
%%CITATION = ARXIV:1211.0083;%%
\bibitem [{\citenamefont {Motohashi}\ \emph {et~al.}(2015)\citenamefont
  {Motohashi}, \citenamefont {Starobinsky},\ and\ \citenamefont
  {Yokoyama}}]{Motohashi:2014ppa}%
  \BibitemOpen
  \bibfield  {author} {\bibinfo {author} {\bibfnamefont {H.}~\bibnamefont
  {Motohashi}}, \bibinfo {author} {\bibfnamefont {A.~A.}\ \bibnamefont
  {Starobinsky}},\ and\ \bibinfo {author} {\bibfnamefont {J.}~\bibnamefont
  {Yokoyama}},\ }\bibfield  {title} {\bibinfo {title} {{Inflation with a
  constant rate of roll}},\ }\href
  {https://doi.org/10.1088/1475-7516/2015/09/018} {\bibfield  {journal}
  {\bibinfo  {journal} {JCAP}\ }\textbf {\bibinfo {volume} {1509}},\ \bibinfo
  {pages} {018}},\ \Eprint {https://arxiv.org/abs/1411.5021} {arXiv:1411.5021
  [astro-ph.CO]} \BibitemShut {NoStop}%
%%CITATION = ARXIV:1411.5021;%%
\bibitem [{\citenamefont {Salopek}\ and\ \citenamefont
  {Stewart}(1992)}]{Salopek:1992qy}%
  \BibitemOpen
  \bibfield  {author} {\bibinfo {author} {\bibfnamefont {D.~S.}\ \bibnamefont
  {Salopek}}\ and\ \bibinfo {author} {\bibfnamefont {J.~M.}\ \bibnamefont
  {Stewart}},\ }\bibfield  {title} {\bibinfo {title} {{Hamilton-Jacobi theory
  for general relativity with matter fields}},\ }\href
  {https://doi.org/10.1088/0264-9381/9/8/015} {\bibfield  {journal} {\bibinfo
  {journal} {Class. Quant. Grav.}\ }\textbf {\bibinfo {volume} {9}},\ \bibinfo
  {pages} {1943} (\bibinfo {year} {1992})}\BibitemShut {NoStop}%
%%CITATION = CQGRD,9,1943;%%
\bibitem [{\citenamefont {Liddle}\ \emph {et~al.}(1994)\citenamefont {Liddle},
  \citenamefont {Parsons},\ and\ \citenamefont {Barrow}}]{Liddle:1994dx}%
  \BibitemOpen
  \bibfield  {author} {\bibinfo {author} {\bibfnamefont {A.~R.}\ \bibnamefont
  {Liddle}}, \bibinfo {author} {\bibfnamefont {P.}~\bibnamefont {Parsons}},\
  and\ \bibinfo {author} {\bibfnamefont {J.~D.}\ \bibnamefont {Barrow}},\
  }\bibfield  {title} {\bibinfo {title} {{Formalizing the slow roll
  approximation in inflation}},\ }\href
  {https://doi.org/10.1103/PhysRevD.50.7222} {\bibfield  {journal} {\bibinfo
  {journal} {Phys. Rev.}\ }\textbf {\bibinfo {volume} {D50}},\ \bibinfo {pages}
  {7222} (\bibinfo {year} {1994})},\ \Eprint
  {https://arxiv.org/abs/astro-ph/9408015} {arXiv:astro-ph/9408015 [astro-ph]}
  \BibitemShut {NoStop}%
%%CITATION = ASTRO-PH/9408015;%%
\bibitem [{\citenamefont {Kinney}(1997)}]{Kinney:1997ne}%
  \BibitemOpen
  \bibfield  {author} {\bibinfo {author} {\bibfnamefont {W.~H.}\ \bibnamefont
  {Kinney}},\ }\bibfield  {title} {\bibinfo {title} {{A Hamilton-Jacobi
  approach to nonslow roll inflation}},\ }\href
  {https://doi.org/10.1103/PhysRevD.56.2002} {\bibfield  {journal} {\bibinfo
  {journal} {Phys. Rev.}\ }\textbf {\bibinfo {volume} {D56}},\ \bibinfo {pages}
  {2002} (\bibinfo {year} {1997})},\ \Eprint
  {https://arxiv.org/abs/hep-ph/9702427} {arXiv:hep-ph/9702427 [hep-ph]}
  \BibitemShut {NoStop}%
%%CITATION = HEP-PH/9702427;%%
\bibitem [{\citenamefont {Guo}\ \emph {et~al.}(2003)\citenamefont {Guo},
  \citenamefont {Piao}, \citenamefont {Cai},\ and\ \citenamefont
  {Zhang}}]{Guo:2003zf}%
  \BibitemOpen
  \bibfield  {author} {\bibinfo {author} {\bibfnamefont {Z.-K.}\ \bibnamefont
  {Guo}}, \bibinfo {author} {\bibfnamefont {Y.-S.}\ \bibnamefont {Piao}},
  \bibinfo {author} {\bibfnamefont {R.-G.}\ \bibnamefont {Cai}},\ and\ \bibinfo
  {author} {\bibfnamefont {Y.-Z.}\ \bibnamefont {Zhang}},\ }\bibfield  {title}
  {\bibinfo {title} {{Inflationary attractor from tachyonic matter}},\ }\href
  {https://doi.org/10.1103/PhysRevD.68.043508} {\bibfield  {journal} {\bibinfo
  {journal} {Phys. Rev.}\ }\textbf {\bibinfo {volume} {D68}},\ \bibinfo {pages}
  {043508} (\bibinfo {year} {2003})},\ \Eprint
  {https://arxiv.org/abs/hep-ph/0304236} {arXiv:hep-ph/0304236 [hep-ph]}
  \BibitemShut {NoStop}%
%%CITATION = HEP-PH/0304236;%%
\bibitem [{\citenamefont {Sheikhahmadi}\ \emph {et~al.}(2016)\citenamefont
  {Sheikhahmadi}, \citenamefont {Saridakis}, \citenamefont {Aghamohammadi},\
  and\ \citenamefont {Saaidi}}]{Sheikhahmadi:2016wyz}%
  \BibitemOpen
  \bibfield  {author} {\bibinfo {author} {\bibfnamefont {H.}~\bibnamefont
  {Sheikhahmadi}}, \bibinfo {author} {\bibfnamefont {E.~N.}\ \bibnamefont
  {Saridakis}}, \bibinfo {author} {\bibfnamefont {A.}~\bibnamefont
  {Aghamohammadi}},\ and\ \bibinfo {author} {\bibfnamefont {K.}~\bibnamefont
  {Saaidi}},\ }\bibfield  {title} {\bibinfo {title} {{Hamilton-Jacobi formalism
  for inflation with non-minimal derivative coupling}},\ }\href
  {https://doi.org/10.1088/1475-7516/2016/10/021} {\bibfield  {journal}
  {\bibinfo  {journal} {JCAP}\ }\textbf {\bibinfo {volume} {1610}}\bibfield
  {number} {\bibinfo  {number} { (10)},\ \bibinfo {pages} {021}},\ }\Eprint
  {https://arxiv.org/abs/1603.03883} {arXiv:1603.03883 [gr-qc]} \BibitemShut
  {NoStop}%
%%CITATION = ARXIV:1603.03883;%%
\bibitem [{\citenamefont {Hsu}(2004)}]{Hsu:2004ri}%
  \BibitemOpen
  \bibfield  {author} {\bibinfo {author} {\bibfnamefont {S.~D.~H.}\
  \bibnamefont {Hsu}},\ }\bibfield  {title} {\bibinfo {title} {{Entropy bounds
  and dark energy}},\ }\href {https://doi.org/10.1016/j.physletb.2004.05.020}
  {\bibfield  {journal} {\bibinfo  {journal} {Phys. Lett. B}\ }\textbf
  {\bibinfo {volume} {594}},\ \bibinfo {pages} {13} (\bibinfo {year} {2004})},\
  \Eprint {https://arxiv.org/abs/hep-th/0403052} {arXiv:hep-th/0403052}
  \BibitemShut {NoStop}%
\bibitem [{\citenamefont {Horvat}(2004)}]{Horvat:2004vn}%
  \BibitemOpen
  \bibfield  {author} {\bibinfo {author} {\bibfnamefont {R.}~\bibnamefont
  {Horvat}},\ }\bibfield  {title} {\bibinfo {title} {{Holography and variable
  cosmological constant}},\ }\href {https://doi.org/10.1103/PhysRevD.70.087301}
  {\bibfield  {journal} {\bibinfo  {journal} {Phys. Rev. D}\ }\textbf {\bibinfo
  {volume} {70}},\ \bibinfo {pages} {087301} (\bibinfo {year} {2004})},\
  \Eprint {https://arxiv.org/abs/astro-ph/0404204} {arXiv:astro-ph/0404204}
  \BibitemShut {NoStop}%
\bibitem [{\citenamefont {Li}(2004)}]{Li:2004rb}%
  \BibitemOpen
  \bibfield  {author} {\bibinfo {author} {\bibfnamefont {M.}~\bibnamefont
  {Li}},\ }\bibfield  {title} {\bibinfo {title} {{A Model of holographic dark
  energy}},\ }\href {https://doi.org/10.1016/j.physletb.2004.10.014} {\bibfield
   {journal} {\bibinfo  {journal} {Phys. Lett. B}\ }\textbf {\bibinfo {volume}
  {603}},\ \bibinfo {pages} {1} (\bibinfo {year} {2004})},\ \Eprint
  {https://arxiv.org/abs/hep-th/0403127} {arXiv:hep-th/0403127} \BibitemShut
  {NoStop}%
\bibitem [{\citenamefont {'t~Hooft}(1993)}]{tHooft:1993dmi}%
  \BibitemOpen
  \bibfield  {author} {\bibinfo {author} {\bibfnamefont {G.}~\bibnamefont
  {'t~Hooft}},\ }\bibfield  {title} {\bibinfo {title} {{Dimensional reduction
  in quantum gravity}},\ }\href@noop {} {\bibfield  {journal} {\bibinfo
  {journal} {Conf. Proc. C}\ }\textbf {\bibinfo {volume} {930308}},\ \bibinfo
  {pages} {284} (\bibinfo {year} {1993})},\ \Eprint
  {https://arxiv.org/abs/gr-qc/9310026} {arXiv:gr-qc/9310026} \BibitemShut
  {NoStop}%
\bibitem [{\citenamefont {Susskind}(1995)}]{Susskind:1994vu}%
  \BibitemOpen
  \bibfield  {author} {\bibinfo {author} {\bibfnamefont {L.}~\bibnamefont
  {Susskind}},\ }\bibfield  {title} {\bibinfo {title} {{The World as a
  hologram}},\ }\href {https://doi.org/10.1063/1.531249} {\bibfield  {journal}
  {\bibinfo  {journal} {J. Math. Phys.}\ }\textbf {\bibinfo {volume} {36}},\
  \bibinfo {pages} {6377} (\bibinfo {year} {1995})},\ \Eprint
  {https://arxiv.org/abs/hep-th/9409089} {arXiv:hep-th/9409089} \BibitemShut
  {NoStop}%
\bibitem [{\citenamefont {Witten}(1998)}]{Witten:1998qj}%
  \BibitemOpen
  \bibfield  {author} {\bibinfo {author} {\bibfnamefont {E.}~\bibnamefont
  {Witten}},\ }\bibfield  {title} {\bibinfo {title} {{Anti-de Sitter space and
  holography}},\ }\href {https://doi.org/10.4310/ATMP.1998.v2.n2.a2} {\bibfield
   {journal} {\bibinfo  {journal} {Adv. Theor. Math. Phys.}\ }\textbf {\bibinfo
  {volume} {2}},\ \bibinfo {pages} {253} (\bibinfo {year} {1998})},\ \Eprint
  {https://arxiv.org/abs/hep-th/9802150} {arXiv:hep-th/9802150} \BibitemShut
  {NoStop}%
\bibitem [{\citenamefont {Bousso}(2002)}]{Bousso:2002ju}%
  \BibitemOpen
  \bibfield  {author} {\bibinfo {author} {\bibfnamefont {R.}~\bibnamefont
  {Bousso}},\ }\bibfield  {title} {\bibinfo {title} {{The Holographic
  principle}},\ }\href {https://doi.org/10.1103/RevModPhys.74.825} {\bibfield
  {journal} {\bibinfo  {journal} {Rev. Mod. Phys.}\ }\textbf {\bibinfo {volume}
  {74}},\ \bibinfo {pages} {825} (\bibinfo {year} {2002})},\ \Eprint
  {https://arxiv.org/abs/hep-th/0203101} {arXiv:hep-th/0203101} \BibitemShut
  {NoStop}%
\bibitem [{\citenamefont {Mann}\ and\ \citenamefont
  {Solodukhin}(1997)}]{Mann:1996ze}%
  \BibitemOpen
  \bibfield  {author} {\bibinfo {author} {\bibfnamefont {R.~B.}\ \bibnamefont
  {Mann}}\ and\ \bibinfo {author} {\bibfnamefont {S.~N.}\ \bibnamefont
  {Solodukhin}},\ }\bibfield  {title} {\bibinfo {title} {{Quantum scalar field
  on three-dimensional (BTZ) black hole instanton: Heat kernel, effective
  action and thermodynamics}},\ }\href
  {https://doi.org/10.1103/PhysRevD.55.3622} {\bibfield  {journal} {\bibinfo
  {journal} {Phys. Rev. D}\ }\textbf {\bibinfo {volume} {55}},\ \bibinfo
  {pages} {3622} (\bibinfo {year} {1997})},\ \Eprint
  {https://arxiv.org/abs/hep-th/9609085} {arXiv:hep-th/9609085} \BibitemShut
  {NoStop}%
\bibitem [{\citenamefont {Rovelli}(1996)}]{Rovelli:1996dv}%
  \BibitemOpen
  \bibfield  {author} {\bibinfo {author} {\bibfnamefont {C.}~\bibnamefont
  {Rovelli}},\ }\bibfield  {title} {\bibinfo {title} {{Black hole entropy from
  loop quantum gravity}},\ }\href {https://doi.org/10.1103/PhysRevLett.77.3288}
  {\bibfield  {journal} {\bibinfo  {journal} {Phys. Rev. Lett.}\ }\textbf
  {\bibinfo {volume} {77}},\ \bibinfo {pages} {3288} (\bibinfo {year}
  {1996})},\ \Eprint {https://arxiv.org/abs/gr-qc/9603063}
  {arXiv:gr-qc/9603063} \BibitemShut {NoStop}%
\bibitem [{\citenamefont {Ashtekar}\ \emph {et~al.}(1998)\citenamefont
  {Ashtekar}, \citenamefont {Baez}, \citenamefont {Corichi},\ and\
  \citenamefont {Krasnov}}]{Ashtekar:1997yu}%
  \BibitemOpen
  \bibfield  {author} {\bibinfo {author} {\bibfnamefont {A.}~\bibnamefont
  {Ashtekar}}, \bibinfo {author} {\bibfnamefont {J.}~\bibnamefont {Baez}},
  \bibinfo {author} {\bibfnamefont {A.}~\bibnamefont {Corichi}},\ and\ \bibinfo
  {author} {\bibfnamefont {K.}~\bibnamefont {Krasnov}},\ }\bibfield  {title}
  {\bibinfo {title} {{Quantum geometry and black hole entropy}},\ }\href
  {https://doi.org/10.1103/PhysRevLett.80.904} {\bibfield  {journal} {\bibinfo
  {journal} {Phys. Rev. Lett.}\ }\textbf {\bibinfo {volume} {80}},\ \bibinfo
  {pages} {904} (\bibinfo {year} {1998})},\ \Eprint
  {https://arxiv.org/abs/gr-qc/9710007} {arXiv:gr-qc/9710007} \BibitemShut
  {NoStop}%
\bibitem [{\citenamefont {Kaul}\ and\ \citenamefont
  {Majumdar}(2000)}]{Kaul:2000kf}%
  \BibitemOpen
  \bibfield  {author} {\bibinfo {author} {\bibfnamefont {R.~K.}\ \bibnamefont
  {Kaul}}\ and\ \bibinfo {author} {\bibfnamefont {P.}~\bibnamefont
  {Majumdar}},\ }\bibfield  {title} {\bibinfo {title} {{Logarithmic correction
  to the Bekenstein-Hawking entropy}},\ }\href
  {https://doi.org/10.1103/PhysRevLett.84.5255} {\bibfield  {journal} {\bibinfo
   {journal} {Phys. Rev. Lett.}\ }\textbf {\bibinfo {volume} {84}},\ \bibinfo
  {pages} {5255} (\bibinfo {year} {2000})},\ \Eprint
  {https://arxiv.org/abs/gr-qc/0002040} {arXiv:gr-qc/0002040} \BibitemShut
  {NoStop}%
\bibitem [{\citenamefont {Das}\ \emph {et~al.}(2002)\citenamefont {Das},
  \citenamefont {Majumdar},\ and\ \citenamefont {Bhaduri}}]{Das:2001ic}%
  \BibitemOpen
  \bibfield  {author} {\bibinfo {author} {\bibfnamefont {S.}~\bibnamefont
  {Das}}, \bibinfo {author} {\bibfnamefont {P.}~\bibnamefont {Majumdar}},\ and\
  \bibinfo {author} {\bibfnamefont {R.~K.}\ \bibnamefont {Bhaduri}},\
  }\bibfield  {title} {\bibinfo {title} {{General logarithmic corrections to
  black hole entropy}},\ }\href {https://doi.org/10.1088/0264-9381/19/9/302}
  {\bibfield  {journal} {\bibinfo  {journal} {Class. Quant. Grav.}\ }\textbf
  {\bibinfo {volume} {19}},\ \bibinfo {pages} {2355} (\bibinfo {year}
  {2002})},\ \Eprint {https://arxiv.org/abs/hep-th/0111001}
  {arXiv:hep-th/0111001} \BibitemShut {NoStop}%
\bibitem [{\citenamefont {Das}\ \emph {et~al.}(2008{\natexlab{a}})\citenamefont
  {Das}, \citenamefont {Shankaranarayanan},\ and\ \citenamefont
  {Sur}}]{Das:2008sy}%
  \BibitemOpen
  \bibfield  {author} {\bibinfo {author} {\bibfnamefont {S.}~\bibnamefont
  {Das}}, \bibinfo {author} {\bibfnamefont {S.}~\bibnamefont
  {Shankaranarayanan}},\ and\ \bibinfo {author} {\bibfnamefont
  {S.}~\bibnamefont {Sur}},\ }\bibfield  {title} {\bibinfo {title} {{Black hole
  entropy from entanglement: A Review}},\ }\href@noop {} {\  (\bibinfo {year}
  {2008}{\natexlab{a}})},\ \Eprint {https://arxiv.org/abs/0806.0402}
  {arXiv:0806.0402 [gr-qc]} \BibitemShut {NoStop}%
\bibitem [{\citenamefont {Das}\ \emph {et~al.}(2010)\citenamefont {Das},
  \citenamefont {Shankaranarayanan},\ and\ \citenamefont {Sur}}]{Das:2010su}%
  \BibitemOpen
  \bibfield  {author} {\bibinfo {author} {\bibfnamefont {S.}~\bibnamefont
  {Das}}, \bibinfo {author} {\bibfnamefont {S.}~\bibnamefont
  {Shankaranarayanan}},\ and\ \bibinfo {author} {\bibfnamefont
  {S.}~\bibnamefont {Sur}},\ }\bibfield  {title} {\bibinfo {title}
  {{Entanglement and corrections to Bekenstein-Hawking entropy}},\ }in\ \href
  {https://doi.org/$10.1142/9789814374552_0152$} {\emph {\bibinfo {booktitle}
  {{12th Marcel Grossmann Meeting on General Relativity}}}}\ (\bibinfo {year}
  {2010})\ pp.\ \bibinfo {pages} {1138--1141},\ \Eprint
  {https://arxiv.org/abs/1002.1129} {arXiv:1002.1129 [gr-qc]} \BibitemShut
  {NoStop}%
\bibitem [{\citenamefont {Das}\ \emph {et~al.}(2008{\natexlab{b}})\citenamefont
  {Das}, \citenamefont {Shankaranarayanan},\ and\ \citenamefont
  {Sur}}]{Das:2007mj}%
  \BibitemOpen
  \bibfield  {author} {\bibinfo {author} {\bibfnamefont {S.}~\bibnamefont
  {Das}}, \bibinfo {author} {\bibfnamefont {S.}~\bibnamefont
  {Shankaranarayanan}},\ and\ \bibinfo {author} {\bibfnamefont
  {S.}~\bibnamefont {Sur}},\ }\bibfield  {title} {\bibinfo {title} {{Power-law
  corrections to entanglement entropy of black holes}},\ }\href
  {https://doi.org/10.1103/PhysRevD.77.064013} {\bibfield  {journal} {\bibinfo
  {journal} {Phys. Rev. D}\ }\textbf {\bibinfo {volume} {77}},\ \bibinfo
  {pages} {064013} (\bibinfo {year} {2008}{\natexlab{b}})},\ \Eprint
  {https://arxiv.org/abs/0705.2070} {arXiv:0705.2070 [gr-qc]} \BibitemShut
  {NoStop}%
\bibitem [{\citenamefont {Radicella}\ and\ \citenamefont
  {Pavon}(2010)}]{Radicella:2010ss}%
  \BibitemOpen
  \bibfield  {author} {\bibinfo {author} {\bibfnamefont {N.}~\bibnamefont
  {Radicella}}\ and\ \bibinfo {author} {\bibfnamefont {D.}~\bibnamefont
  {Pavon}},\ }\bibfield  {title} {\bibinfo {title} {{The generalized second law
  in universes with quantum corrected entropy relations}},\ }\href
  {https://doi.org/10.1016/j.physletb.2010.06.019} {\bibfield  {journal}
  {\bibinfo  {journal} {Phys. Lett. B}\ }\textbf {\bibinfo {volume} {691}},\
  \bibinfo {pages} {121} (\bibinfo {year} {2010})},\ \Eprint
  {https://arxiv.org/abs/1006.3745} {arXiv:1006.3745 [gr-qc]} \BibitemShut
  {NoStop}%
\bibitem [{\citenamefont {Nojiri}\ \emph
  {et~al.}(2019{\natexlab{a}})\citenamefont {Nojiri}, \citenamefont
  {Odintsov},\ and\ \citenamefont {Saridakis}}]{Nojiri:2019skr}%
  \BibitemOpen
  \bibfield  {author} {\bibinfo {author} {\bibfnamefont {S.}~\bibnamefont
  {Nojiri}}, \bibinfo {author} {\bibfnamefont {S.~D.}\ \bibnamefont
  {Odintsov}},\ and\ \bibinfo {author} {\bibfnamefont {E.~N.}\ \bibnamefont
  {Saridakis}},\ }\bibfield  {title} {\bibinfo {title} {{Modified cosmology
  from extended entropy with varying exponent}},\ }\href
  {https://doi.org/10.1140/epjc/s10052-019-6740-5} {\bibfield  {journal}
  {\bibinfo  {journal} {Eur. Phys. J. C}\ }\textbf {\bibinfo {volume} {79}},\
  \bibinfo {pages} {242} (\bibinfo {year} {2019}{\natexlab{a}})},\ \Eprint
  {https://arxiv.org/abs/1903.03098} {arXiv:1903.03098 [gr-qc]} \BibitemShut
  {NoStop}%
\bibitem [{\citenamefont {Nojiri}\ \emph
  {et~al.}(2020{\natexlab{a}})\citenamefont {Nojiri}, \citenamefont {Odintsov},
  \citenamefont {Saridakis},\ and\ \citenamefont
  {Myrzakulov}}]{Nojiri:2019itp}%
  \BibitemOpen
  \bibfield  {author} {\bibinfo {author} {\bibfnamefont {S.}~\bibnamefont
  {Nojiri}}, \bibinfo {author} {\bibfnamefont {S.~D.}\ \bibnamefont
  {Odintsov}}, \bibinfo {author} {\bibfnamefont {E.~N.}\ \bibnamefont
  {Saridakis}},\ and\ \bibinfo {author} {\bibfnamefont {R.}~\bibnamefont
  {Myrzakulov}},\ }\bibfield  {title} {\bibinfo {title} {{Correspondence of
  cosmology from non-extensive thermodynamics with fluids of generalized
  equation of state}},\ }\href
  {https://doi.org/10.1016/j.nuclphysb.2019.114850} {\bibfield  {journal}
  {\bibinfo  {journal} {Nucl. Phys. B}\ }\textbf {\bibinfo {volume} {950}},\
  \bibinfo {pages} {114850} (\bibinfo {year} {2020}{\natexlab{a}})},\ \Eprint
  {https://arxiv.org/abs/1911.03606} {arXiv:1911.03606 [gr-qc]} \BibitemShut
  {NoStop}%
\bibitem [{\citenamefont {Nojiri}\ and\ \citenamefont
  {Odintsov}(2006)}]{Nojiri:2005pu}%
  \BibitemOpen
  \bibfield  {author} {\bibinfo {author} {\bibfnamefont {S.}~\bibnamefont
  {Nojiri}}\ and\ \bibinfo {author} {\bibfnamefont {S.~D.}\ \bibnamefont
  {Odintsov}},\ }\bibfield  {title} {\bibinfo {title} {{Unifying phantom
  inflation with late-time acceleration: Scalar phantom-non-phantom transition
  model and generalized holographic dark energy}},\ }\href
  {https://doi.org/10.1007/s10714-006-0301-6} {\bibfield  {journal} {\bibinfo
  {journal} {Gen. Rel. Grav.}\ }\textbf {\bibinfo {volume} {38}},\ \bibinfo
  {pages} {1285} (\bibinfo {year} {2006})},\ \Eprint
  {https://arxiv.org/abs/hep-th/0506212} {arXiv:hep-th/0506212} \BibitemShut
  {NoStop}%
\bibitem [{\citenamefont {Nojiri}\ \emph
  {et~al.}(2020{\natexlab{b}})\citenamefont {Nojiri}, \citenamefont {Odintsov},
  \citenamefont {Oikonomou},\ and\ \citenamefont {Paul}}]{Nojiri:2020wmh}%
  \BibitemOpen
  \bibfield  {author} {\bibinfo {author} {\bibfnamefont {S.}~\bibnamefont
  {Nojiri}}, \bibinfo {author} {\bibfnamefont {S.}~\bibnamefont {Odintsov}},
  \bibinfo {author} {\bibfnamefont {V.}~\bibnamefont {Oikonomou}},\ and\
  \bibinfo {author} {\bibfnamefont {T.}~\bibnamefont {Paul}},\ }\bibfield
  {title} {\bibinfo {title} {{Unifying Holographic Inflation with Holographic
  Dark Energy: a Covariant Approach}},\ }\href
  {https://doi.org/10.1103/PhysRevD.102.023540} {\bibfield  {journal} {\bibinfo
   {journal} {Phys. Rev. D}\ }\textbf {\bibinfo {volume} {102}},\ \bibinfo
  {pages} {023540} (\bibinfo {year} {2020}{\natexlab{b}})},\ \Eprint
  {https://arxiv.org/abs/2007.06829} {arXiv:2007.06829 [gr-qc]} \BibitemShut
  {NoStop}%
\bibitem [{\citenamefont {Nojiri}\ \emph
  {et~al.}(2019{\natexlab{b}})\citenamefont {Nojiri}, \citenamefont
  {Odintsov},\ and\ \citenamefont {Saridakis}}]{Nojiri:2019kkp}%
  \BibitemOpen
  \bibfield  {author} {\bibinfo {author} {\bibfnamefont {S.}~\bibnamefont
  {Nojiri}}, \bibinfo {author} {\bibfnamefont {S.~D.}\ \bibnamefont
  {Odintsov}},\ and\ \bibinfo {author} {\bibfnamefont {E.~N.}\ \bibnamefont
  {Saridakis}},\ }\bibfield  {title} {\bibinfo {title} {{Holographic
  inflation}},\ }\href {https://doi.org/10.1016/j.physletb.2019.134829}
  {\bibfield  {journal} {\bibinfo  {journal} {Phys. Lett. B}\ }\textbf
  {\bibinfo {volume} {797}},\ \bibinfo {pages} {134829} (\bibinfo {year}
  {2019}{\natexlab{b}})},\ \Eprint {https://arxiv.org/abs/1904.01345}
  {arXiv:1904.01345 [gr-qc]} \BibitemShut {NoStop}%
\bibitem [{\citenamefont {Oliveros}\ and\ \citenamefont
  {Acero}(2019)}]{Oliveros:2019rnq}%
  \BibitemOpen
  \bibfield  {author} {\bibinfo {author} {\bibfnamefont {A.}~\bibnamefont
  {Oliveros}}\ and\ \bibinfo {author} {\bibfnamefont {M.~A.}\ \bibnamefont
  {Acero}},\ }\bibfield  {title} {\bibinfo {title} {{Inflation driven by a
  holographic energy density}},\ }\href
  {https://doi.org/10.1209/0295-5075/128/59001} {\bibfield  {journal} {\bibinfo
   {journal} {EPL}\ }\textbf {\bibinfo {volume} {128}},\ \bibinfo {pages}
  {59001} (\bibinfo {year} {2019})},\ \Eprint
  {https://arxiv.org/abs/1911.04482} {arXiv:1911.04482 [gr-qc]} \BibitemShut
  {NoStop}%
\bibitem [{\citenamefont {Chakraborty}\ and\ \citenamefont
  {Chattopadhyay}(2020)}]{Chakraborty_2020}%
  \BibitemOpen
  \bibfield  {author} {\bibinfo {author} {\bibfnamefont {G.}~\bibnamefont
  {Chakraborty}}\ and\ \bibinfo {author} {\bibfnamefont {S.}~\bibnamefont
  {Chattopadhyay}},\ }\bibfield  {title} {\bibinfo {title} {Modified
  holographic energy density-driven inflation and some cosmological outcomes},\
  }\href {https://doi.org/10.1142/s0219887820500668} {\bibfield  {journal}
  {\bibinfo  {journal} {International Journal of Geometric Methods in Modern
  Physics}\ }\textbf {\bibinfo {volume} {17}},\ \bibinfo {pages} {2050066}
  (\bibinfo {year} {2020})}\BibitemShut {NoStop}%
\bibitem [{\citenamefont {Mohammadi}\ \emph {et~al.}(2021)\citenamefont
  {Mohammadi}, \citenamefont {Golanbari}, \citenamefont {Bamba},\ and\
  \citenamefont {Lobo}}]{Mohammadi:2021wde}%
  \BibitemOpen
  \bibfield  {author} {\bibinfo {author} {\bibfnamefont {A.}~\bibnamefont
  {Mohammadi}}, \bibinfo {author} {\bibfnamefont {T.}~\bibnamefont
  {Golanbari}}, \bibinfo {author} {\bibfnamefont {K.}~\bibnamefont {Bamba}},\
  and\ \bibinfo {author} {\bibfnamefont {I.~P.}\ \bibnamefont {Lobo}},\
  }\bibfield  {title} {\bibinfo {title} {{Tsallis holographic dark energy for
  inflation}},\ }\href {https://doi.org/10.1103/PhysRevD.103.083505} {\bibfield
   {journal} {\bibinfo  {journal} {Phys. Rev. D}\ }\textbf {\bibinfo {volume}
  {103}},\ \bibinfo {pages} {083505} (\bibinfo {year} {2021})},\ \Eprint
  {https://arxiv.org/abs/2101.06378} {arXiv:2101.06378 [gr-qc]} \BibitemShut
  {NoStop}%
\bibitem [{\citenamefont {Motohashi}\ and\ \citenamefont
  {Starobinsky}(2017)}]{Motohashi:2017aob}%
  \BibitemOpen
  \bibfield  {author} {\bibinfo {author} {\bibfnamefont {H.}~\bibnamefont
  {Motohashi}}\ and\ \bibinfo {author} {\bibfnamefont {A.~A.}\ \bibnamefont
  {Starobinsky}},\ }\bibfield  {title} {\bibinfo {title} {{Constant-roll
  inflation: confrontation with recent observational data}},\ }\href
  {https://doi.org/10.1209/0295-5075/117/39001} {\bibfield  {journal} {\bibinfo
   {journal} {EPL}\ }\textbf {\bibinfo {volume} {117}},\ \bibinfo {pages}
  {39001} (\bibinfo {year} {2017})},\ \Eprint
  {https://arxiv.org/abs/1702.05847} {arXiv:1702.05847 [astro-ph.CO]}
  \BibitemShut {NoStop}%
\bibitem [{\citenamefont {Motohashi}\ and\ \citenamefont
  {Starobinsky}(2019)}]{Motohashi:2019tyj}%
  \BibitemOpen
  \bibfield  {author} {\bibinfo {author} {\bibfnamefont {H.}~\bibnamefont
  {Motohashi}}\ and\ \bibinfo {author} {\bibfnamefont {A.~A.}\ \bibnamefont
  {Starobinsky}},\ }\bibfield  {title} {\bibinfo {title} {{Constant-roll
  inflation in scalar-tensor gravity}},\ }\href
  {https://doi.org/10.1088/1475-7516/2019/11/025} {\bibfield  {journal}
  {\bibinfo  {journal} {JCAP}\ }\textbf {\bibinfo {volume} {11}},\ \bibinfo
  {pages} {025}},\ \Eprint {https://arxiv.org/abs/1909.10883} {arXiv:1909.10883
  [gr-qc]} \BibitemShut {NoStop}%
\bibitem [{\citenamefont {Motohashi}\ \emph {et~al.}(2020)\citenamefont
  {Motohashi}, \citenamefont {Mukohyama},\ and\ \citenamefont
  {Oliosi}}]{Motohashi:2019rhu}%
  \BibitemOpen
  \bibfield  {author} {\bibinfo {author} {\bibfnamefont {H.}~\bibnamefont
  {Motohashi}}, \bibinfo {author} {\bibfnamefont {S.}~\bibnamefont
  {Mukohyama}},\ and\ \bibinfo {author} {\bibfnamefont {M.}~\bibnamefont
  {Oliosi}},\ }\bibfield  {title} {\bibinfo {title} {{Constant Roll and
  Primordial Black Holes}},\ }\href
  {https://doi.org/10.1088/1475-7516/2020/03/002} {\bibfield  {journal}
  {\bibinfo  {journal} {JCAP}\ }\textbf {\bibinfo {volume} {03}},\ \bibinfo
  {pages} {002}},\ \Eprint {https://arxiv.org/abs/1910.13235} {arXiv:1910.13235
  [gr-qc]} \BibitemShut {NoStop}%
\bibitem [{\citenamefont {Granda}\ and\ \citenamefont
  {Oliveros}(2008)}]{Granda:2008dk}%
  \BibitemOpen
  \bibfield  {author} {\bibinfo {author} {\bibfnamefont {L.}~\bibnamefont
  {Granda}}\ and\ \bibinfo {author} {\bibfnamefont {A.}~\bibnamefont
  {Oliveros}},\ }\bibfield  {title} {\bibinfo {title} {{Infrared cut-off
  proposal for the Holographic density}},\ }\href
  {https://doi.org/10.1016/j.physletb.2008.10.017} {\bibfield  {journal}
  {\bibinfo  {journal} {Phys. Lett. B}\ }\textbf {\bibinfo {volume} {669}},\
  \bibinfo {pages} {275} (\bibinfo {year} {2008})},\ \Eprint
  {https://arxiv.org/abs/0810.3149} {arXiv:0810.3149 [gr-qc]} \BibitemShut
  {NoStop}%
\bibitem [{\citenamefont {Granda}\ and\ \citenamefont
  {Oliveros}(2009)}]{Granda:2008tm}%
  \BibitemOpen
  \bibfield  {author} {\bibinfo {author} {\bibfnamefont {L.}~\bibnamefont
  {Granda}}\ and\ \bibinfo {author} {\bibfnamefont {A.}~\bibnamefont
  {Oliveros}},\ }\bibfield  {title} {\bibinfo {title} {{New infrared cut-off
  for the holographic scalar fields models of dark energy}},\ }\href
  {https://doi.org/10.1016/j.physletb.2008.12.025} {\bibfield  {journal}
  {\bibinfo  {journal} {Phys. Lett. B}\ }\textbf {\bibinfo {volume} {671}},\
  \bibinfo {pages} {199} (\bibinfo {year} {2009})},\ \Eprint
  {https://arxiv.org/abs/0810.3663} {arXiv:0810.3663 [gr-qc]} \BibitemShut
  {NoStop}%
\bibitem [{\citenamefont {Radicella}\ and\ \citenamefont
  {Pav{\'{o}}n}(2010)}]{Radicella_2010}%
  \BibitemOpen
  \bibfield  {author} {\bibinfo {author} {\bibfnamefont {N.}~\bibnamefont
  {Radicella}}\ and\ \bibinfo {author} {\bibfnamefont {D.}~\bibnamefont
  {Pav{\'{o}}n}},\ }\bibfield  {title} {\bibinfo {title} {On thec2term in the
  holographic formula for dark energy},\ }\href
  {https://doi.org/10.1088/1475-7516/2010/10/005} {\bibfield  {journal}
  {\bibinfo  {journal} {Journal of Cosmology and Astroparticle Physics}\
  }\textbf {\bibinfo {volume} {2010}}\bibinfo  {number} { (10)},\ \bibinfo
  {pages} {005}}\BibitemShut {NoStop}%
\bibitem [{\citenamefont {Pavon}\ and\ \citenamefont
  {Zimdahl}(2005)}]{Pavon:2005yx}%
  \BibitemOpen
\bibfield  {number} {  }\bibfield  {author} {\bibinfo {author} {\bibfnamefont
  {D.}~\bibnamefont {Pavon}}\ and\ \bibinfo {author} {\bibfnamefont
  {W.}~\bibnamefont {Zimdahl}},\ }\bibfield  {title} {\bibinfo {title}
  {{Holographic dark energy and cosmic coincidence}},\ }\href
  {https://doi.org/10.1016/j.physletb.2005.08.134} {\bibfield  {journal}
  {\bibinfo  {journal} {Phys. Lett. B}\ }\textbf {\bibinfo {volume} {628}},\
  \bibinfo {pages} {206} (\bibinfo {year} {2005})},\ \Eprint
  {https://arxiv.org/abs/gr-qc/0505020} {arXiv:gr-qc/0505020} \BibitemShut
  {NoStop}%
\bibitem [{\citenamefont {Pav{\'{o}}n}(2007)}]{Pav_n_2007}%
  \BibitemOpen
  \bibfield  {author} {\bibinfo {author} {\bibfnamefont {D.}~\bibnamefont
  {Pav{\'{o}}n}},\ }\bibfield  {title} {\bibinfo {title} {Holographic dark
  energy and late cosmic acceleration},\ }\href
  {https://doi.org/10.1088/1751-8113/40/25/s31} {\bibfield  {journal} {\bibinfo
   {journal} {Journal of Physics A: Mathematical and Theoretical}\ }\textbf
  {\bibinfo {volume} {40}},\ \bibinfo {pages} {6865} (\bibinfo {year}
  {2007})}\BibitemShut {NoStop}%
\bibitem [{\citenamefont {Guberina}\ \emph {et~al.}(2007)\citenamefont
  {Guberina}, \citenamefont {Horvat},\ and\ \citenamefont
  {Nikoli{\'{c}}}}]{Guberina_2007}%
  \BibitemOpen
  \bibfield  {author} {\bibinfo {author} {\bibfnamefont {B.}~\bibnamefont
  {Guberina}}, \bibinfo {author} {\bibfnamefont {R.}~\bibnamefont {Horvat}},\
  and\ \bibinfo {author} {\bibfnamefont {H.}~\bibnamefont {Nikoli{\'{c}}}},\
  }\bibfield  {title} {\bibinfo {title} {Non-saturated holographic dark
  energy},\ }\href {https://doi.org/10.1088/1475-7516/2007/01/012} {\bibfield
  {journal} {\bibinfo  {journal} {Journal of Cosmology and Astroparticle
  Physics}\ }\textbf {\bibinfo {volume} {2007}}\bibinfo  {number} { (01)},\
  \bibinfo {pages} {012}}\BibitemShut {NoStop}%
\bibitem [{\citenamefont {Malekjani}\ \emph {et~al.}(2018)\citenamefont
  {Malekjani}, \citenamefont {Rezaei},\ and\ \citenamefont
  {Akhlaghi}}]{Malekjani_prd_2018}%
  \BibitemOpen
\bibfield  {number} {  }\bibfield  {author} {\bibinfo {author} {\bibfnamefont
  {M.}~\bibnamefont {Malekjani}}, \bibinfo {author} {\bibfnamefont
  {M.}~\bibnamefont {Rezaei}},\ and\ \bibinfo {author} {\bibfnamefont {I.~A.}\
  \bibnamefont {Akhlaghi}},\ }\bibfield  {title} {\bibinfo {title} {Can
  holographic dark energy models fit the observational data?},\ }\href
  {https://doi.org/10.1103/PhysRevD.98.063533} {\bibfield  {journal} {\bibinfo
  {journal} {Phys. Rev. D}\ }\textbf {\bibinfo {volume} {98}},\ \bibinfo
  {pages} {063533} (\bibinfo {year} {2018})}\BibitemShut {NoStop}%
\bibitem [{\citenamefont {Ade}\ \emph {et~al.}(2021)\citenamefont {Ade} \emph
  {et~al.}}]{BICEP:2021xfz}%
  \BibitemOpen
  \bibfield  {author} {\bibinfo {author} {\bibfnamefont {P.~A.~R.}\
  \bibnamefont {Ade}} \emph {et~al.} (\bibinfo {collaboration} {BICEP, Keck}),\
  }\bibfield  {title} {\bibinfo {title} {{Improved Constraints on Primordial
  Gravitational Waves using Planck, WMAP, and BICEP/Keck Observations through
  the 2018 Observing Season}},\ }\href
  {https://doi.org/10.1103/PhysRevLett.127.151301} {\bibfield  {journal}
  {\bibinfo  {journal} {Phys. Rev. Lett.}\ }\textbf {\bibinfo {volume} {127}},\
  \bibinfo {pages} {151301} (\bibinfo {year} {2021})},\ \Eprint
  {https://arxiv.org/abs/2110.00483} {arXiv:2110.00483 [astro-ph.CO]}
  \BibitemShut {NoStop}%
\bibitem [{\citenamefont {Chattopadhyay}\ \emph {et~al.}(2014)\citenamefont
  {Chattopadhyay}, \citenamefont {Pasqua},\ and\ \citenamefont
  {Khurshudyan}}]{Chattopadhyay:2014yda}%
  \BibitemOpen
  \bibfield  {author} {\bibinfo {author} {\bibfnamefont {S.}~\bibnamefont
  {Chattopadhyay}}, \bibinfo {author} {\bibfnamefont {A.}~\bibnamefont
  {Pasqua}},\ and\ \bibinfo {author} {\bibfnamefont {M.}~\bibnamefont
  {Khurshudyan}},\ }\bibfield  {title} {\bibinfo {title} {{New holographic
  reconstruction of scalar field dark energy models in the framework of
  chameleon Brans-Dicke cosmology}},\ }\href
  {https://doi.org/10.1140/epjc/s10052-014-3080-3} {\bibfield  {journal}
  {\bibinfo  {journal} {Eur. Phys. J. C}\ }\textbf {\bibinfo {volume} {74}},\
  \bibinfo {pages} {3080} (\bibinfo {year} {2014})},\ \Eprint
  {https://arxiv.org/abs/1401.8208} {arXiv:1401.8208 [gr-qc]} \BibitemShut
  {NoStop}%
\bibitem [{\citenamefont {Chattopadhyay}\ and\ \citenamefont
  {Pasqua}(2016)}]{Chattopadhyay:2016enn}%
  \BibitemOpen
  \bibfield  {author} {\bibinfo {author} {\bibfnamefont {S.}~\bibnamefont
  {Chattopadhyay}}\ and\ \bibinfo {author} {\bibfnamefont {A.}~\bibnamefont
  {Pasqua}},\ }\bibfield  {title} {\bibinfo {title} {{Consequences of
  Holographic Scalar Field Dark Energy Models in Chameleon Brans-Dicke
  Cosmology}},\ }\href {https://doi.org/$10.1007/978-3-319-25619-1_74$}
  {\bibfield  {journal} {\bibinfo  {journal} {Springer Proc. Phys.}\ }\textbf
  {\bibinfo {volume} {174}},\ \bibinfo {pages} {487} (\bibinfo {year}
  {2016})}\BibitemShut {NoStop}%
\bibitem [{\citenamefont {Samanta}(2013)}]{Samanta:2013vna}%
  \BibitemOpen
  \bibfield  {author} {\bibinfo {author} {\bibfnamefont {G.~C.}\ \bibnamefont
  {Samanta}},\ }\bibfield  {title} {\bibinfo {title} {{Holographic Dark Energy
  (DE) Cosmological Models with Quintessence in Bianchi Type-V Space Time}},\
  }\href {https://doi.org/10.1007/s10773-013-1757-2} {\bibfield  {journal}
  {\bibinfo  {journal} {Int. J. Theor. Phys.}\ }\textbf {\bibinfo {volume}
  {52}},\ \bibinfo {pages} {4389} (\bibinfo {year} {2013})}\BibitemShut
  {NoStop}%
\bibitem [{\citenamefont {Sheykhi}(2011)}]{Sheykhi:2011}%
  \BibitemOpen
  \bibfield  {author} {\bibinfo {author} {\bibfnamefont {A.}~\bibnamefont
  {Sheykhi}},\ }\bibfield  {title} {\bibinfo {title} {Holographic scalar field
  models of dark energy},\ }\bibfield  {journal} {\bibinfo  {journal} {Physical
  Review D}\ }\textbf {\bibinfo {volume} {84}},\ \href
  {https://doi.org/10.1103/physrevd.84.107302} {10.1103/physrevd.84.107302}
  (\bibinfo {year} {2011})\BibitemShut {NoStop}%
\bibitem [{\citenamefont {Yang}\ \emph {et~al.}(2011)\citenamefont {Yang},
  \citenamefont {Wu}, \citenamefont {Song}, \citenamefont {Su}, \citenamefont
  {Li}, \citenamefont {Zhang},\ and\ \citenamefont {Wang}}]{Yang:2011zza}%
  \BibitemOpen
  \bibfield  {author} {\bibinfo {author} {\bibfnamefont {W.-Q.}\ \bibnamefont
  {Yang}}, \bibinfo {author} {\bibfnamefont {Y.-B.}\ \bibnamefont {Wu}},
  \bibinfo {author} {\bibfnamefont {L.-M.}\ \bibnamefont {Song}}, \bibinfo
  {author} {\bibfnamefont {Y.-Y.}\ \bibnamefont {Su}}, \bibinfo {author}
  {\bibfnamefont {J.}~\bibnamefont {Li}}, \bibinfo {author} {\bibfnamefont
  {D.-D.}\ \bibnamefont {Zhang}},\ and\ \bibinfo {author} {\bibfnamefont
  {X.-G.}\ \bibnamefont {Wang}},\ }\bibfield  {title} {\bibinfo {title}
  {{Reconstruction of new holographic scalar field models of dark energy in
  Brans-Dicke universe}},\ }\href {https://doi.org/10.1142/S0217732311034682}
  {\bibfield  {journal} {\bibinfo  {journal} {Mod. Phys. Lett. A}\ }\textbf
  {\bibinfo {volume} {26}},\ \bibinfo {pages} {191} (\bibinfo {year} {2011})},\
  \Eprint {https://arxiv.org/abs/1311.5884} {arXiv:1311.5884 [gr-qc]}
  \BibitemShut {NoStop}%
\bibitem [{\citenamefont {Karami}\ and\ \citenamefont
  {Fehri}(2010)}]{Karami:2009we}%
  \BibitemOpen
  \bibfield  {author} {\bibinfo {author} {\bibfnamefont {K.}~\bibnamefont
  {Karami}}\ and\ \bibinfo {author} {\bibfnamefont {J.}~\bibnamefont {Fehri}},\
  }\bibfield  {title} {\bibinfo {title} {{New holographic scalar field models
  of dark energy in non-flat universe}},\ }\href
  {https://doi.org/10.1016/j.physletb.2009.12.060} {\bibfield  {journal}
  {\bibinfo  {journal} {Phys. Lett. B}\ }\textbf {\bibinfo {volume} {684}},\
  \bibinfo {pages} {61} (\bibinfo {year} {2010})},\ \Eprint
  {https://arxiv.org/abs/0912.1541} {arXiv:0912.1541 [gr-qc]} \BibitemShut
  {NoStop}%
\end{thebibliography}%

%\begin{thebibliography}
%\bibitem{Rasheed:2020syk} A.~Mohammadi, T.~Golanbari, H.~Sheikhahmadi,
%K.~Sayar, L.~Akhtari, M.~A.~Rasheed and K.~Saaidi, ``Warm Tachyon Inflation
%and Swampland Criteria,'' Chinese Physics C \textbf{\ 44}, No. 9 (2020)
%095101, doi:10.1088/1674-1137/44/9/095101 [arXiv:2001.10042 [gr-qc]].
%2 citations counted in INSPIRE as of 11 Jul 2020
%\end{thebibliography}

\end{document}